\documentstyle[prb,aps,epsfig,a4]{revtex}
\begin{document}
\title{\bf Measurements of the parameters of the $\phi(1020)$ resonance 
           through studies of the processes \\
	   $e^+e^- \to K^+K^-$, $K_SK_L$ and $\pi^+\pi^-\pi^0$}   
\author{ M.N.Achasov\thanks{ E-mail: achasov@inp.nsk.su, 
FAX: +7(383-2)34-21-63},
         K.I.Beloborodov, A.V.Berdyugin,A.G.Bogdanchikov,
	 A.V.Bozhenok, A.D.Bukin, D.A.Bukin, S.V.Burdin,
	 T.V.Dimova, V.P.Druzhinin, M.S.Dubrovin, I.A.Gaponenko,
	 V.B.Golubev, V.N.Ivanchenko, P.M.Ivanov, A.A.Korol,
	 M.S.Korostelev, S.V.Koshuba, E.V.Pakhtusova, A.A.Polunin,
	 E.E.Pyata, A.A.Salnikov, S.I.Serednyakov,  V.V.Shary,
	 Yu.M.Shatunov, V.A.Sidorov, Z.K.Silagadze,
	 A.N.Skrinsky, Yu.V.Usov, A.V.Vasiljev, Yu.S.Velikzhanin }
\address{ 
          Budker Institute of Nuclear Physics,  \\
          Siberian Branch of the Russian Academy of Sciences and \\
	  Novosibirsk State University, \\
          11 Laurentyev,Novosibirsk, \\
	  630090, Russia}          

\maketitle

\begin{abstract}
 The cross sections of the processes $e^+e^- \to K^+K^-$, $e^+e^- \to K_SK_L$
 and $e^+e^- \to \pi^+\pi^-\pi^0$ were measured in the SND experiment at the
 VEPP-2M collider in the energy region near the $\phi(1020)$ meson. These
 measurements were based on about $10^6$ $K^+K^-$, $0.5 \times 10^6$ $K_SK_L$
 and $0.4 \times 10^6$ $\pi^+\pi^-\pi^0$ selected events. The
 measured cross sections have been analyzed in the framework of the vector
 meson dominance model and the main parameters of the $\phi$-resonance were
 obtained, such as its mass, width, the production cross section and branching
 ratios of the main decay modes. The measured value of the  $\phi$ meson total
 width, $\Gamma_{\phi} = 4.21 \pm 0.04$ is lower than the present world average
 of $4.458 \pm 0.032$ MeV. Contributions in addition to the conventional vector
 meson dominance model were found in the $e^+e^- \to \pi^+\pi^-\pi^0$
 reaction cross section.
\end{abstract}

\section{Introduction}

 Cross sections of the reactions $e^+e^- \to K\overline{K}$ and
 $e^+e^- \to \pi^+\pi^-\pi^0$, are determined by strong and
 electromagnetic interactions of the light quarks $(s,u,d)$ and cannot be
 evaluated at present from first principles. But a rather good description
 of these cross sections is provided by the vector meson dominance  model
 (VDM) with phenomenologically adjusted coupling constants
 ($g_{V\gamma}$, $g_{V\rho\pi}$, $g_{VP\gamma}$, $g_{VPP}$). In the VDM, the
 cross sections are determined by the amplitudes of vector meson V transitions
 ($V = \phi$, $\omega$, $\rho$) into the final state:
 $V \to \rho \pi \to 3\pi$, $V \to K\overline{K}$. In the energy range from
 980 to 1070 MeV, the main contributions to the $e^+e^- \to K\overline{K}$ and
 $e^+e^- \to \pi^+\pi^-\pi^0$ cross sections come from the $\phi$-meson.
 Therefore the measurement of these cross sections allows us to determine the
 $\phi$-resonance main parameters and study its interference with other
 vector mesons. Such studies provide important information about light meson
 physics.
 
 Earlier these processes were studied in several experiments:
 $e^+e^- \to 3\pi$ \cite{orse1,olia,orse2,nd,cmd1,cmd2}, $e^+e^- \to K_SK_L$
 \cite{nd,cmd1,olia1,cmd3} and $e^+e^- \to K^+K^-$ \cite{cmd1,olia2,ospk}.
 The SND (Spherical Neutral Detector) experiment continues these efforts at
 the VEPP-2M collider  \cite{vepp2}. Here we present the results of this
 investigation.

\section{Experiment}

 The SND has been operating at the VEPP-2M
 collider in the energy range from 360 to 1400 MeV since 1995 (the SND detailed
 description can be found in Ref.\cite{sndnim}). The detector contains several
 subsystems. The tracking system includes two cylindrical drift chambers. The
 three-layer spherical electromagnetic calorimeter is based on NaI(Tl)
 crystals. The muon/veto system consists of plastic scintillation counters and
 two layers of streamer tubes. The calorimeter energy and
 angular resolution depend on the photon energy as
 $\sigma_E/E(\%) = {4.2\% / \sqrt[4]{E(\mbox{GeV})}}$ and
 $\sigma_{\phi,\theta} = {0.82^\circ / \sqrt[]{E(\mathrm{GeV})}}
 \oplus 0.63^\circ$.
 The tracking system angular resolution is about $0.5^\circ$ and $2^\circ$ for
 azimuthal and polar angle respectively. The energy loss resolution 
 $dE/dx$ in the drift chamber is about 30\% -- good enough to
 provide charged kaon identification in the region of $\phi$-meson production.

 This work is based on the data sample collected in the SND experiments during
 1998. The data were collected at 32 energy points in the energy
 range from 984 to 1060 MeV (Table \ref{tab1}). The total integrated
 luminosity accumulated in these experiments is 8.5 pb$^{-1}$. The beam
 energy was calculated from the magnetic field value in the bending magnets
 and revolution frequency of the collider. In the vicinity of the $\phi$-meson
 peak we also used the beam energy calibrated by the Cryogenic Magnetic
 Detector (CMD-2), which operated at VEPP-2M at the same time. This
 calibration \cite{cmd3,cmd4} is based on the measurements of the charged
 kaon momenta in the drift chamber of CMD-2. The energy calibration accuracy
 in each point is about 0.1 MeV, the common shift of the energy scale is
 estimated to be 0.04 MeV, and the energy spread of the beams is about 0.37
 MeV.

 For the luminosity measurements, $e^+e^- \to e^+e^-$ and 
 $e^+e^- \to \gamma\gamma$ processes were used. The corresponding luminosity 
 values were respectively used for normalizing the events containing charged
 particles and those with neutral particles only. They are different by about
 10\% (Tab.\ref{tab1}) due to the dead time of the trigger, which selected
 events containing neutral particles only. The systematic error of the
 integrated luminosity determination is estimated to be 2\%. Since luminosity
 measurements by  $e^+e^- \to e^+e^-$ and $e^+e^- \to \gamma\gamma$ reveal a
 systematic spread of about 1.3\%, this  was added to the statistical error
 of the luminosity determination in each  energy point. The statistical
 accuracy was better than 1\%.
 
 The experimental conditions were rather stable during data taking:
 the SND systems counting rate, average currents and luminosity of the collider
 were stable at the level of about 20 \% and no correlation between different
 energy points was observed.

\section{Data analysis}

 The procedure of the experimental data selection consists of several 
 stages. During the experimental runs, the first-level trigger was used
 \cite{sndnim}. The first-level trigger selects events of different types:
 events with photons only and events with charged particles. The thresholds
 applied on the energy deposition in the calorimeter were about 200 MeV
 for events containing charged particles and 300 MeV for events with neutral
 particles only.
 
 During the processing of experimental data, the event reconstruction is
 performed in the following sequence. The first step is a search for separated
 clusters in the calorimeter. Track reconstruction in the drift chambers is
 then performed. Tracks are linked to the calorimeter clusters. Clusters with
 energy depositions of more than 20 MeV, and not linked to tracks in the drift
 chambers are considered as photons. For charged tracks no requirements on
 the energy depositions in the calorimeter are imposed.

 For further analysis, events with $|z| < 10$ cm and $r < 1$ cm for each 
 charged particle were selected. Here $z$ is the coordinate of the charged
 particle production point along the beam axis (the longitudinal size of the
 interaction region is $\sigma_z \sim 2.5$ cm); $r$ is the distance between a
 charged particle track and the beam axis in the $r-\phi$ plane. For each
 process individual selection criteria were then applied.

 The detection efficiency was obtained from Monte-Carlo (MC) simulation.
 Simulated events were reconstructed and the actual selection criteria were
 applied. To take into account the apparatus resolution of the first-level
 trigger thresholds, for simulated and experimental events the trigger
 conditions were applied with the threshold values 10\% higher than during
 the experimental runs.

 As a result of uncertainties in the simulation of the distributions over some
 selection parameters, the detection efficiency was multiplied by correction
 coefficients, which were obtained in the following way. The experimental
 events were selected without any conditions on the parameter under study,
 using the cut parameters uncorrelated with the studied one.
 The same selection was applied for simulated events. Then the cut was
 applied to the parameter and the correction coefficient was calculated:
\begin{eqnarray}
\label{poprawka}
 \delta = {{n/N} \over {m/M}},
\end{eqnarray}
 where $N$ and $M$ are the number of events in experiment and simulation
 selected without any cuts on the parameter under study; $n$ and $m$ is
 the number of events in experiment and simulation, when the cut on the
 parameter was applied. As a rule, the error in the coefficient  $\delta$
 determination is connected with the uncertainty of background subtraction.
 This systematic error was estimated by varying other selection criteria.

 The overlap of the beam background with the events containing charged
 particles can result in track reconstruction failure and a decrease of
 detection efficiency. To take into account this effect, background events
 (experimental events collected when detector was triggered with an external
 generator) were overlapped with the simulated events. It was found that the
 detection efficiency decreased by about 3\% and therefore the correction
 coefficient $\delta_{over} = 0.97 \pm 0.03$ was used for the processes
 containing charged particles. In the analysis described further simulated
 events are not overlapped with beam background events.

\subsection{Selection of $e^+e^- \to K^+K^-$ events}

\begin{figure}[h]
\begin{center}
\epsfig{figure=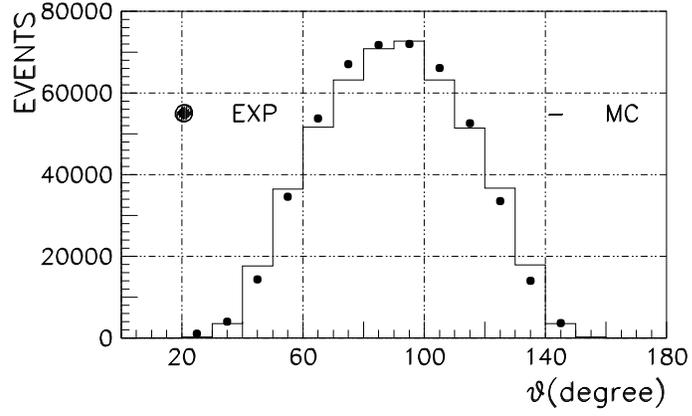,width=9cm}
\caption{The $\theta$ distribution of charged particles from
the $e^+e^-\to K^+K^-$ reaction.}
\label{kkc_tet}
\end{center}
\end{figure}

 In the studied energy region the $e^+e^- \to K^+K^-$ events have the 
 following  features:
\begin{itemize}
\item
 The charged kaons are rather slow ($\beta\gamma \sim 0.2$ -- $0.4$) and
 therefore have large $dE/dx$ losses in the drift chamber.
\item
 Events often contain several photons due to the kaon decays
 ($K^\pm \to \pi^\pm\pi^0$, $\pi^\pm\pi^0\pi^0$,$\pi^0\mu^\pm\nu_\mu$,
 $\pi^0 e^\pm\nu_e$) in the tracking system.
\item
 Tracks attributed to the kaon decays in the tracking system
 (secondary particles) are also present.
\end{itemize}
 To select the $e^+e^- \to K^+K^-$ events,  detection
 of two charged particles was required, with polar angles in the range
 $20^{\circ} < \theta < 160^{\circ}$, with acollinearity angle in the azimuthal
 plane $\Delta\phi < 10^\circ$ and the energy losses in the drift chamber
 $dE/dx > 3 \cdot (dE/dx)_{min}$, where $(dE/dx)_{min}$ is an average $(dE/dx)$
 loss of a minimum ionizing particle. The events must also contain two or
 more photons and at least one secondary particle from kaon decay
 in the tracking system. The distribution of the angle $\theta$ for
 charged particles in the selected events has the typical shape
 for the $e^+e^-\to K^+K^-$ reaction (Fig.\ref{kkc_tet}).
 
 After these cuts, the background processes   $e^+e^- \to \pi^+\pi^-$,
 $\mu^+\mu^-$ and also beam and cosmic backgrounds still contribute. 
 The data accumulated under the $K\overline{K}$ production 
 threshold, at the energy point
 $\sqrt[]{s} = 984$ MeV, were used to estimate the number of background events.
 After selection criteria were applied, only $N_{bkg}(984) = 5$ events were
 found at this point. The number of background events in each energy point was
 calculated as
\begin{eqnarray}
 N_{bkg}({s}) = N_{bkg}(984) \cdot { IL({s}) \over IL(984)},
\label{nfon}
\end{eqnarray}
 where $IL({s})$ is the integrated luminosity, and this number was subtracted
 from the number of selected events.

 The detection efficiency was multiplied by correction coefficients
 $\delta_{dE/dx}$, $\delta_{over}$, $\delta_{NP}$, $\delta_{etot}$. Here
 $\delta_{dE/dx}=0.93\pm0.02$ is a correction due to the inaccuracy in the 
 simulation of the $dE/dx$ distributions. This coefficient does not depend on
 the energy in the explored region. The coefficient $\delta_{NP}$ is a 
 correction to the efficiency of detection of secondary particles and at least
 two photons.
 Its value varies in this energy range by $\sim$ 10\% and in the
 vicinity of the $\phi$-resonance peak is $\delta_{NP} = 0.99 \pm 0.015$. 
 The correction $\delta_{etot}=0.91 \pm 0.05$ is due to the inaccuracy in the
 simulation of the average energy deposition of  $e^+e^- \to K^+K^-$ events.
 The total systematic error of the detection efficiency determination is
 6.8\%. The number of selected $e^+e^- \to K^+K^-$ events (after background 
 subtraction) and the detection
 efficiency are shown in Table~\ref{tab2}. 
 The detection efficiency rises with energy because the probability of  
 kaon absorption in the detector passive material decreases with energy.
 Above 1035 MeV the efficiency goes down since the probability of kaon decay
 in the  detector decreases.

\subsection{Selection of $e^+e^- \to K_S K_L$ events}

 The $K_S$ and $K_L$ mesons decay lengths in the studied energy region
 are $c\tau\beta\gamma \sim 0.3$ -- $1$ cm and $2$ -- $5.6$ m respectively.
 The $K_S$ meson decays inside the collider vacuum chamber or tracking system,
 and the $K_L$ reaches the calorimeter, where it can either produce signals
 (``photons'') due to nuclear interactions or decay in flight, or it can pass
 through the detector without interacting. The analysis of the
 $e^+e^- \to K_SK_L$ reaction is based on the detection of
 $K_S \to \pi^+\pi^-$ or $K_S \to \pi^0\pi^0$ decays. 

 In the first case (charged mode) the events which contained two charged
 particles
 and at least one ``photon'' were used.  The polar angles of the charged
 particles and  their acollinearity angle were bounded by the criteria:
 $36^{\circ} < \theta < 144^{\circ}$, $|\Delta\theta| < 30^\circ$ and
 $10^\circ < |\Delta\phi| < 50^\circ$. The cut $10^\circ < |\Delta\phi|$
 rejects the background from $e^+e^- \to K^+K^-$ events. To suppress 
 the same  background, the following cuts were also applied:
 $(dE/dx) < 5 \cdot (dE/dx)_{min}$ for each charged particle and 
 $(dE/dx) < 3 \cdot (dE/dx)_{min} $ at least for one.

\begin{figure}[h]
\begin{center}
\epsfig{figure=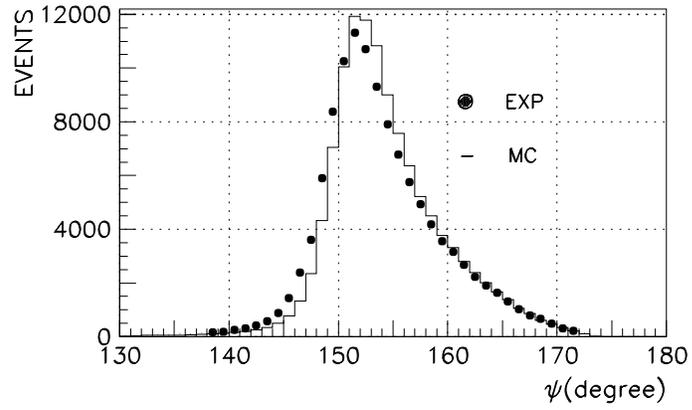,width=9cm}
\caption{Angle $\psi$ between pion pairs from the $K_S \to \pi^+\pi^-$ decay.}
\label{kslan12}
\end{center}
\end{figure}

 The $K_S$ meson decay length is about 0.5 cm, so to suppress the
 background from $e^+e^- \to 3\pi$,
 $\eta\gamma(\eta \to \pi^+\pi^-\pi^0, \pi^+\pi^- \gamma)$, $\omega\pi^0$,
 $e^+e^-\gamma\gamma$ events, the cut $0.2$ cm $< r < 1$ was used.
 The beam and cosmic backgrounds were suppressed by the criteria: $E_{p} > 0$
 and $E_{p3} < 50 $ MeV, where $E_{p}$ and $E_{p3}$ are the total energy
 deposition and the energy deposition in the third calorimeter layer of
 any charged particle . Pions in the process under study have energies of about
 $\sim 200$ -- $300$ MeV and actually give a very low
 energy deposition in the third layer. The distribution of the angle $\psi$
 between charged particles in the selected events is shown in
 Figure~\ref{kslan12}. It has a characteristic peak at the minimum angle 
 between pions,  close to $150^\circ$.

 To estimate the number of background events, the data accumulated
 at the energy point ~$\sqrt[]{s} = 984$ MeV, i.e. under the reaction
 threshold, were used. After the cuts, $N_{bkg}(984) = 30$ events were left.
 Their production
 points are uniformly distributed along the beam direction, so these events 
 can
 be attributed to the beam and cosmic backgrounds. The number of background
 events in other energy points was calculated according to the
 formula (\ref{nfon}) and then subtracted from the number of selected events.

\begin{figure}[h]
\begin{center}
\epsfig{figure=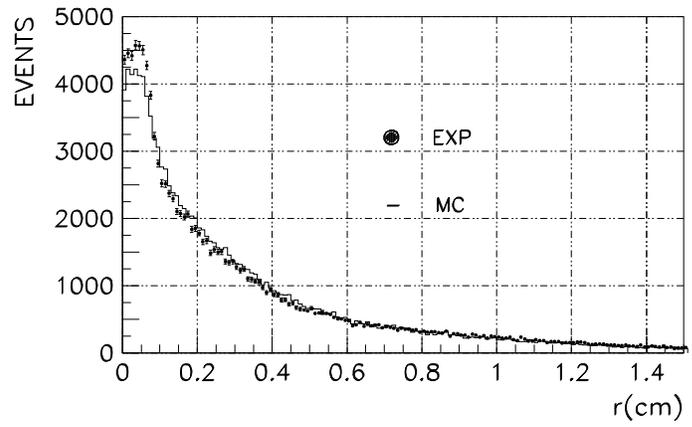,width=9cm}
\caption{The $r$ distribution of  pions from the $K_S \to \pi^+\pi^-$ decay}
\label{kslr}
\end{center}
\end{figure}

 The detection efficiency was multiplied by correction coefficients
 $\delta_{dE/dx}$, $\delta_{over}$ and $\delta_{r}$. Here
 $\delta_{dE/dx} = 0.95 \pm 0.01$ is a correction due to an inaccuracy in the
 simulation of the $dE/dx$ distribution. The coefficient $\delta_{r}$ is a
 correction due to an inaccuracy in the simulation of the $r$ distribution
 (Fig.\ref{kslr}). Its value varies in this energy region by $\sim$ 10\% and in
 the vicinity of the $\phi$-resonance peak is $\delta_r = 0.91 \pm 0.015$.
 The total systematic error of the detection efficiency determination is
 3.7\%. The numbers of selected $e^+e^- \to K_SK_L$ events (after background 
 subtraction) and the detection efficiencies are shown in Table~\ref{tab21}.
 The detection efficiency depends on the energy due to the dependence of
 $\Delta\phi$ and $\Delta \theta$ distributions on the $K_S$ energy and the
 dependence of $K_L$ nuclear interactions and decay length on its energy.

 For analysis of the $e^+e^- \to K_SK_L$ decay in the neutral mode
 ($K_S \to \pi^0\pi^0$), we used events where at least four 
 photons were detected and no charged particles were present. To
 reject beam and cosmic backgrounds, constraints were imposed on the total
 energy deposition ($E_{tot}>0.35\cdot\sqrt[]{s}$) and total momentum of events
 ($P_{tot}<0.6\cdot\sqrt[]{s}$). To suppress cosmic background even further,
 the events where most of the hit crystals could be fitted by a single 
 track, were rejected.

\begin{figure}[h]
\begin{center}
\epsfig{figure=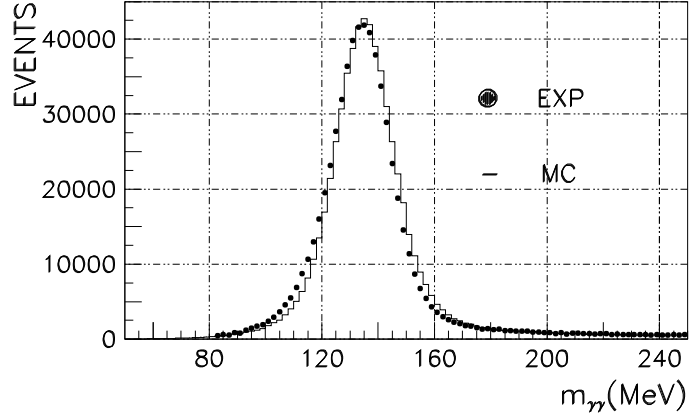,width=9cm}
\caption{The invariant mass of photon pairs from the 
$K_S \to \pi^0\pi^0$ decay.}
\label{ksl_mpi}
\end{center}
\end{figure}

\begin{figure}[h]
\begin{center}
\epsfig{figure=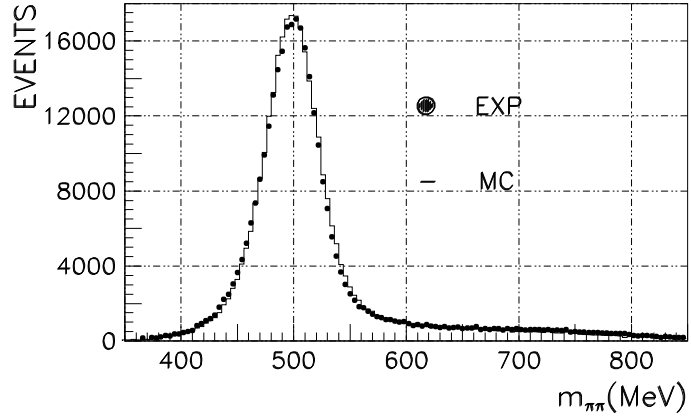,width=9cm}
\caption{The invariant mass of  pions from the 
$K_S \to \pi^0\pi^0$
         decay}
\label{ksl_mks}
\end{center}
\end{figure}

\begin{figure}[h]
\begin{center}
\epsfig{figure=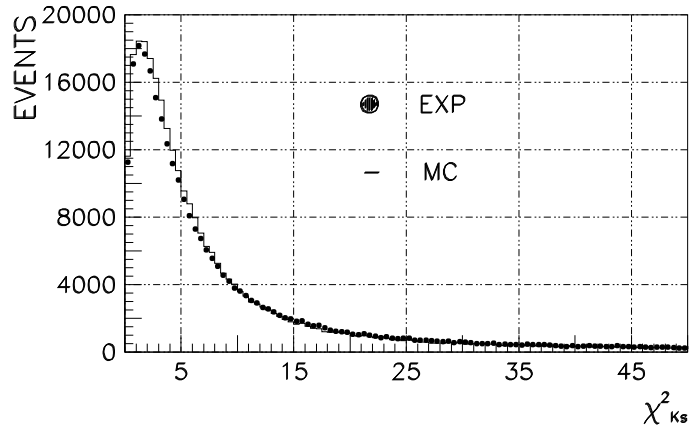,width=9cm}
\caption{The $\chi^2_{K_S}$ distribution for the $K_S \to \pi^0\pi^0$ events.}
\label{ksl_chi2}
\end{center}
\end{figure}

\begin{figure}[h]
\begin{center}
\epsfig{figure=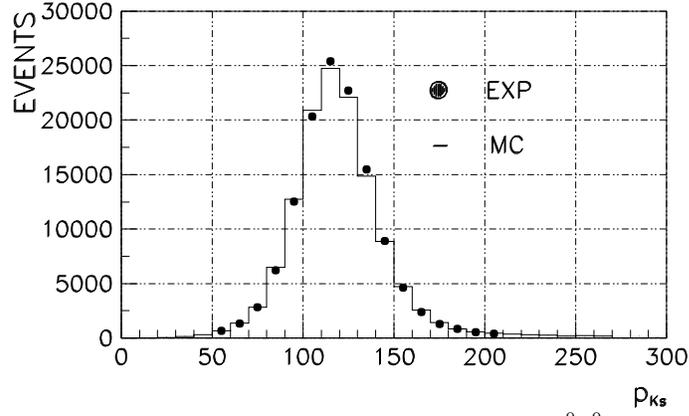,width=9cm}
\caption{The kaon momentum distribution from the $K_S \to \pi^0\pi^0$
         decay at the energy $\sqrt[]{s}=1020$ MeV.}
\label{ksl_mo1}
\end{center}
\end{figure}

\begin{figure}[h]
\begin{center}
\epsfig{figure=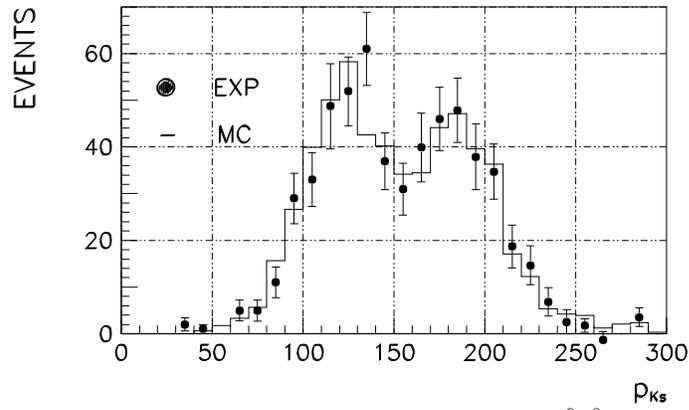,width=9cm}
\caption{The kaon momentum distribution from the $K_S \to \pi^0\pi^0$
         decay at the energy $\sqrt[]{s}=1060$ MeV. The left peak corresponds
	 to the case when the photon was emitted by the initial particles.}
\label{ksl_mo2}
\end{center}
\end{figure}

 For events satisfying selection criteria described above, the kinematic fit
 was performed under the following assumptions: the event contains two photon
 pairs originating from the $\pi^0 \to \gamma\gamma$ decay and invariant 
 mass of
 these four photons is equal to the $K_S$-meson mass. The value of the
 likelihood function $\chi^2_{K_S}$ is calculated. In events with more than
 four photons, extra photons are considered as spurious ones and rejected. To
 do this, all possible subsets of four photons were tested and one, 
 corresponding to the maximum likelihood, was selected.
 Figures \ref{ksl_mpi} and \ref{ksl_mks} represent invariant masses of
 photon pairs and of all four photons found in the reconstruction.
 To suppress the beam background, polar angles of the photons,
 reconstructed as $\gamma$-quanta from the $K_S \to 2\pi^0 \to 4\gamma$ decay,
 were limited to $40^\circ < \theta < 140^\circ$. The background from the
 multi-photon events (pure QED events, $e^+e^-\to\pi^0\pi^0\gamma$
 and $\eta\gamma$) was rejected by the following cuts:~$\chi^2_{K_S}<20$ and
 $\Delta p<100$~MeV. Here $\Delta p = p_{K_S}-\sqrt[]{s/4-m_{K_S}^2}$ and
 $p_{K_S}$ is the measured kaon momentum. The $\chi^2_{K_S}$ and
 $p_{K_S}$ distributions are shown in Figures~\ref{ksl_chi2},~\ref{ksl_mo1} and
 \ref{ksl_mo2}.

 After these cuts were applied, only three events were found at the energy point
 ~$\sqrt[]{s} = 984$ MeV. The number of background events for each energy
 point was calculated according to Eq.(\ref{nfon}) and then subtracted from the
 number of selected events. The number of background events
 from the $e^+e^- \to \eta \gamma$ decay was estimated using simulation and 
 even in the vicinity of the $\phi$-resonance peak, their number is less than
 0.2 \% of the selected $e^+e^- \to K_SK_L$ events.

 The detection efficiency was multiplied by correction coefficients:
 $\delta_{\theta}$ and $\delta_{\chi^2_{K_S}}$.
 The $\delta_{\chi^2_{K_S}} = 0.97 \pm 0.01$ is a correction due to an 
 inaccuracy in the simulation of the $\chi^2_{K_S}$ distribution. The
 experimental $e^+e^- \to K_SK_L$ events can contain false photons, caused by
 the beam background overlaps. The energy of these photons is similar to the
 energy of the $\gamma$-quanta from the $K_S$ decay, and they are localized at
 the polar angles close to the beam direction. 
 During reconstruction, some false photons can be mistaken as originating from 
 the $K_S$ decay. In other words, the reconstruction program can reject with 
 some probability a valid photon from the $K_S$ decay and substitute it 
 by the false one. An event reconstructed in such a way will not pass our cut
 on the photon polar angles (described above) because the beam related false
 photons have small polar angles. Therefore the beam background overlaps lead 
 to the decrease of the detection efficiency -- an effect not taken into
 account in the MC simulation. Hence the necessity of the correction
 coefficient  $\delta_{\theta} = 0.90 \pm 0.03$ arises.
 The detection efficiency does not depend on the energy and equals
 0.068 after all corrections were implemented. The total systematic error of
 the detection efficiency determination is 3.5\%. The selected event numbers
 (after background subtraction) are shown in Table~\ref{tab22}.

\subsection{Selection of $e^+e^- \to \pi^+\pi^-\pi^0$ events}

 For analysis of the $e^+e^- \to \pi^+\pi^-\pi^0$ process, events containing
 two charged particles and two or more photons were used. Extra photons
 can appear because of the beam background overlap, splitting of the 
 electromagnetic showers, and nuclear interactions of the charged pions in the
 calorimeter. Under these selection conditions, the background sources are 
 $e^+e^- \to K^+K^-$, $K_SK_L(K_S \to \pi^+\pi^-)$,
 $\eta\gamma(\eta \to \pi^+\pi^-\pi^0, \pi^+\pi^- \gamma)$, $\omega\pi^0$,
 $e^+e^-\gamma\gamma$ processes and the beam background.

 The polar angles of the charged particles were bounded by the criterion:
 $20^{\circ} < \theta < 160^{\circ}$. To suppress the beam
 background,
 the following cuts on the spatial angle between two charged particles and 
 energy deposition of the neutral particles were applied:
 $\psi > 40^{\circ}$, $E_{neu} > 0.1 \cdot \sqrt[]{s}$.

 To suppress the background from the $e^+e^- \to K^+K^-$ process, the following
 cuts were imposed: $(dE/dx) < 5 \cdot (dE/dx)_{min} $ for each charged
 particle, $(dE/dx) < 3 \cdot (dE/dx)_{min} $ at least for one and
 $\Delta\phi < 10^\circ$. To reject the  $e^+e^- \to e^+e^-\gamma\gamma$
 events the energy deposition of the charged particles was bounded by the
 criterion: $E_{cha} < 0.5 \cdot \sqrt[]{s}$.

\begin{figure}
\begin{center}
\epsfig{figure=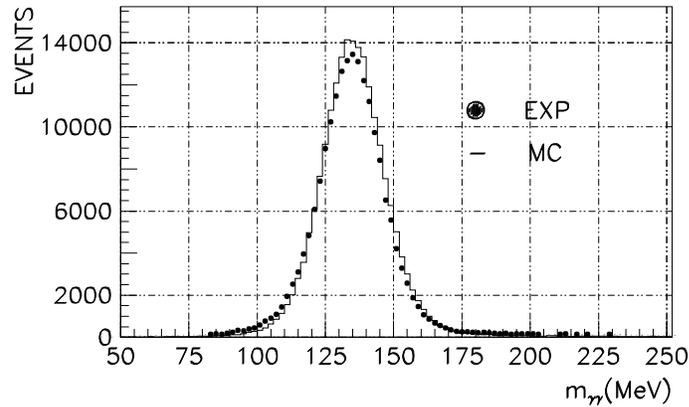,width=9cm}
\caption{Two-photon invariant mass distribution in the 
         $\phi \rightarrow \pi^+ \pi^- \pi^0$ events.}
\label{mgg}
\end{center}
\end{figure}

\begin{figure}
\begin{center}
\epsfig{figure=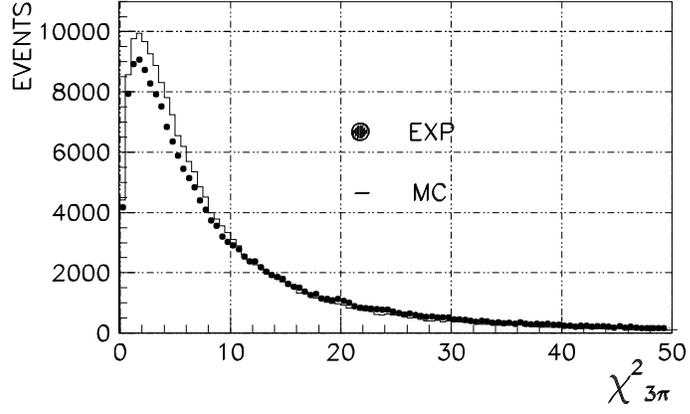,width=9cm}
\caption{The $\chi^2_{3\pi}$ distribution in $e^+e^- \to \pi^+\pi^-\pi^0$
         events.}
\label{chi}
\end{center}
\end{figure}

 For events left after these cuts, a kinematic fit was performed under the
 following constraints: the charged particles are considered to be pions, the
 system has zero momentum, the total energy is $\sqrt[]{s}$, and the photons
 originate from the
 $\pi^0 \to \gamma\gamma$ decays. The value of the likelihood function 
 $\chi^2_{3\pi}$ is calculated. In events with more than two photons,
 extra photons are considered as spurious ones and rejected. To do this, all
 possible subsets of two photons were tested and the one, corresponding to the
 maximum likelihood was selected. Two-photon invariant mass and
 $\chi^2_{3\pi}$ distributions are shown in Figures \ref{mgg} and
 \ref{chi}. After the kinematic fit, the following additional cuts 
 were applied:
 $36^{\circ} < \theta_\gamma < 144^{\circ}$,~~$NNP=2$, and $\chi^2_{3\pi} < 20$
 for energy points with ~~$\sqrt[]{s}<1028$ MeV, and $\chi^2_{3\pi} < 5$
 for energy points above 1028 MeV. Here $\theta_\gamma$ is polar angle of
 any photon selected by the reconstruction program as originated from the
 $\pi^0$-decay. $NNP$ is the number of detected photons.
 A harder cut on $\chi^2_{3\pi}$ at the five last energy points is
 necessary to suppress $e^+e^- \to 3\pi \gamma_{rad}$ events, where
 $\gamma_{rad}$ is a photon emitted by initial particles.

\begin{figure}[h]
\begin{center}
\epsfig{figure=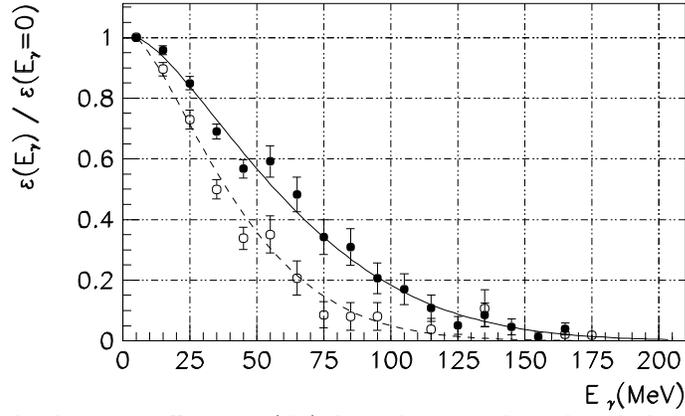,width=9cm}
\caption{The detection efficiency $\epsilon(E_\gamma)$ dependence on the
         radiated photon energy $E_\gamma$ for $e^+e^- \to 3\pi+\gamma$
	 events under the conditions: $\chi^2_{3\pi} < 20$ (dots) and
	 $\chi^2_{3\pi} < 5$ (circles), obtained by simulation.}
\label{pi3rad}
\end{center}
\end{figure}

 As experimental data were accumulated in a rather narrow energy region,
 the detection efficiency for the events without $\gamma$-quantum radiation
 by initial particles does not depend on energy in the entire region.
 The detection efficiency dependence on the radiated photon energy,
 obtained from simulation, is shown in Figure~\ref{pi3rad}.

 The detection efficiency was multiplied by correction coefficients:
$\delta_{over}$, $\delta_{dE/dx} = 0.95 \pm 0.01$,
$\delta_{\chi^2_{3\pi}} = 0.93 \pm 0.03$ and $\delta_{NNP}=0.87 \pm 0.005$.
The last correction factor is due to inaccuracies in the simulation of extra
 photons.
 The detection efficiency (without $\gamma$-quantum radiation) equals 0.183 
 (under condition the $\chi^2_{3\pi}<20$) and 0.086 (under condition the
 $\chi^2_{3\pi}<5$) after all corrections were implemented.  The total
 systematic error of the detection efficiency determination is 4.5\%. The
 selected event numbers are shown in Table~\ref{tab22}.

 The number of background events ($e^+e^- \to \omega \pi^0$, $\eta \gamma$,
 $e^+e^-\gamma\gamma$ and $K \overline{K}$) was estimated using simulation
 in the following way:
\begin{eqnarray}
 N_{bkg}({s}) = \sum_i \sigma_{Ri}({s}) \epsilon_i({s}) \delta_i IL({s}), 
\end{eqnarray}
 where $i$ is a background process number, $\sigma_{Ri}({s})$ is
 the cross section of the process taking into account the radiative
 corrections, $IL({s})$ is the integrated luminosity,
 $\epsilon_i({s})$ is the detection efficiency and $\delta_i$ is a correction
 to the efficiency. For $e^+e^- \to K \overline{K}$ reactions the
 $\sigma_{Ri}({s})$ were taken according to
 the SND measurements presented in this
 article and for the process $e^+e^- \to \omega \pi^0$ other SND measurements
 \cite{vd} were used. The $e^+e^- \to K_SK_L$ is the main background in the
 vicinity of the $\phi$-resonance peak ($\sqrt[]{s} = 1015.75$ -- $1028.23$
 MeV) and its contribution is  $\sim$ 90\% of all background. Outside this
 region the background is strongly determined by the $e^+e^- \to \omega \pi^0$
 and $e^+e^-\gamma\gamma$ events. The background from the  $e^+e^- \to K^+K^-$ 
 events is negligible, and even at the $\phi$-meson peak it is only
 $\sim 0.5 \%$ of all background events. The accuracy of background event
 numbers determination is less than 15\% at all energy points. The obtained
 number $N_{bkg}({s})$ was subtracted from the selected events. The numbers of
 the $e^+e^- \to 3\pi$ and background events are shown in Table~\ref{tab3}.

\section{Theoretical cross sections}

 In the VDM framework the cross sections of the
 $e^+e^- \to K^+K^-$, $K_SK_L$
 and $\pi^+\pi^-\pi^0$ process are written as follows \cite{frlook1}:
  
\begin{eqnarray}
 \sigma_{K^+K^-}(s) = { {8\pi\alpha} \over {3 s^{5/2}} } q^3_{K^+K^-}(s) \cdot
 \Biggl| {{g_{\phi\gamma} g_{\phi K^+K^-}} \over {D_{\phi}(s)} }
 e^{i\phi_{K\overline{K}}}
 - {{g_{\omega\gamma} g_{\omega K^+K^-}} \over {D_{\omega}(s)} } - \nonumber \\
 - {{g_{\rho\gamma} g_{\rho K^+K^-}} \over {D_{\rho}(s)} }
 + A_{K^+K^-} \Biggr|^2
\end{eqnarray}     
\begin{eqnarray}
 \sigma_{K_SK_L}(s) = { {8\pi\alpha} \over {3 s^{5/2}} } q^3_{K_SK_L}(s) \cdot
 \Biggl| {{g_{\phi\gamma} g_{\phi K_SK_L}} \over {D_{\phi}(s)} }
 e^{i\phi_{K\overline{K}}}
 - {{g_{\omega\gamma} g_{\omega K_SK_L}} \over {D_{\omega}(s)} } + \nonumber \\
 + {{g_{\rho\gamma} g_{\rho K_SK_L}} \over {D_{\rho}(s)} }
 + A_{K_SK_L} \Biggr|^2
\end{eqnarray}
\begin{eqnarray}
\label{pi3sig0}
 \sigma_{3\pi}(s) = { {4\pi\alpha} \over {s^{3/2}} } W(s) \cdot
 \Biggl| {{g_{\phi\gamma} g_{\phi\rho\pi}} \over {D_{\phi}(s)} }
 e^{i\chi_{\phi-\omega}} +
 {{g_{\omega\gamma} g_{\omega\rho\pi}} \over {D_{\omega}(s)} } + A_{3\pi}
 \Biggr|^2,
 \mbox{where}
\end{eqnarray}
\begin{eqnarray}
 D_{V} = m_{V}^2 - s -
 i\sqrt{s}\Gamma_{V}(s), \nonumber \\
 \Gamma_{V}(s) = \Gamma(V \to 3\pi,s) + \Gamma(V \to \eta \gamma,s) +
 \Gamma(V \to \pi^0 \gamma,s) + \nonumber \\
 + \Gamma(V \to K^+K^-,s) +  \Gamma(V \to K_SK_L,s) +
 \Gamma(V \to \pi^+ \pi^-,s).
\end{eqnarray}
 Here $V$  denotes the vector mesons $\rho,\omega,\phi$; $q_{K^+K^-}(s)$,
 $q_{K_SK_L}(s)$ are kaon momenta, $W(s)$ is a phase space factor including
 $\rho\pi$ intermediate state \cite{frlook1}, $A_{K^+K^-}$,
 $A_{K_SK_L}$ and $A_{3\pi}$ are amplitudes corresponding to contributions
 additional to the conventional VDM (for example $\rho$, $\omega$ and $\phi$
 primes), and $\chi_{\phi-\omega}$ and $\phi_{K\overline{K}}$ 
 are the relative interference phases. The $\phi_{K\overline{K}}$ and
 $\chi_{\phi-\omega}$ phases are equal to $180^\circ$ in the framework of the
 naive quark model.
 
 The partial decay width dependences on energy are written as:
\begin{eqnarray}
 \Gamma(V \to 3\pi,s) = {{|g_{V\rho\pi}|^2}\over{4\pi}}W(s), \mbox{~~}
 \Gamma(V \to P P,s) = {{|g_{VPP}|^2}\over{6\pi s}}q^3_{PP}(s),
\nonumber \\
 \Gamma(V \to P \gamma,s) = {1 \over 3} |g_{VP \gamma}|^2
 q^3_{P \gamma}(s),
\end{eqnarray}
 where $P$ denotes the pseudoscalar meson $\pi$ or $K$.

 The coupling constants are determined through the decay branching ratios
in the following way:
\begin{eqnarray}
\label{g}
 |g_{V\gamma}| = \Biggl[ {{3m_V^3\Gamma_VB(V \to e^+e^-)} \over
   {4\pi\alpha}} \Biggr]^{1/2}, \mbox{~~}
 |g_{VP\gamma}| = \Biggl[{{3\Gamma_VB(V \to P\gamma)} \over
 {q^3_{P \gamma}(m_V)}} \Biggr]^{1/2},
\nonumber \\
 |g_{V\rho\pi}| = \Biggl[{{4\pi\Gamma_VB(V \to \rho\pi)} \over
 {W(m_V)}} \Biggr]^{1/2}, \mbox{~~}
 |g_{VPP}| = \Biggl[{{6\pi m_V^2\Gamma_VB(V \to PP)} \over
 {q^3_{PP}(m_V)}} \Biggr]^{1/2},
\nonumber \\
 |g_{\omega (\rho) K^+K^-}| = {{1} \over {\sqrt{2}}} |g_{\phi K^+K^-}|,
 \mbox{~~}
 |g_{\omega (\rho) K_SK_L}| = {{1} \over {\sqrt{2}}} |g_{\phi K_SK_L}| 
\end{eqnarray}
 Here $\Gamma_V = \Gamma_V(m_V)$. To restrict the growth of the
 partial width $\Gamma(\omega \to 3\pi, s)$ with energy, 
 the form factor is usually included 
 \cite{frlook1} in the $g_{\omega\rho\pi}$ coupling 
 constant:
 $g_{\omega\rho\pi} \to g_{\omega\rho\pi}\sqrt{C_{\omega\rho\pi}(s)}$, where
\begin{eqnarray}
 C_{\omega\rho\pi}(s) = \Biggl[{{1+(R_{\omega\rho\pi}m_\omega)^2} \over
 {1+(R_{\omega\rho\pi}\sqrt{s})^2}} \Biggr]^2
\end{eqnarray}
 We did not use this form factor for the experimental data approximation in 
this work: the parameter $R_{\omega\rho\pi}$ was set to zero.
 
 The Coulomb interaction of the charged kaons
 in the final state was taken into account by multiplying the coupling 
constant $g_{\phi K^+K^-}$ by the appropriate
 form factor:
 $g_{\phi K^+K^-} \to g_{\phi K^+K^-} \cdot  \sqrt{Z(s)/Z(m_\phi)}$,
 where   \cite{kform}:
\begin{eqnarray}
 Z(s) = 1 + \alpha\pi {{1+v^2} \over {2v}}, \mbox{~~}
 v = \Biggl(1-{ {4m_{K^\pm}^2} \over {s} } \Biggr)^{1/2}
\nonumber
\end{eqnarray}
 In the vicinity of the $\phi$-meson peak, the value $Z(s)$ is nearly
 constant
 and the influence of the $\sqrt{Z(s)/Z(m_\phi)}$ form factor on the resonance
 width and mass is negligible.

 For the experimental data approximation, theoretical cross sections were 
 reduced to the following expressions:
\begin{eqnarray}
\label{kkcsigf}
 \sigma_{K^+K^-}(s) = { {1} \over {s^{5/2}} }
 {{q^3_{K^+K^-}(s)} \over {q^3_{K^+K^-}(m_\phi)}} \cdot
 \Biggl| { {\Gamma_\phi m_\phi^3 \sqrt{\sigma(\phi\to K^+K^-)m_\phi}} \over
 {D_{\phi}(s)} } e^{i\phi_{K\overline{K}}} 
\nonumber \\
 - { {\sqrt{\Gamma_\phi\Gamma_\omega m_\omega^3 m_\phi^2 6\pi
 B(\omega\to e^+e^-)B(\phi\to K^+K^-)}} \over {D_{\omega}(s)} } \\
 - { {\sqrt{\Gamma_\phi\Gamma_\rho m_\rho^3 m_\phi^2 6\pi
 B(\rho\to e^+e^-)B(\phi\to K^+K^-)}} \over {D_{\rho}(s)} }
 + A^0_{K^+K^-}\Biggr|^2 ;
\nonumber
\end{eqnarray}
\begin{eqnarray}
\label{kslsigf}
 \sigma_{K_SK_L}(s) = { {1} \over {s^{5/2}} }
 {{q^3_{K_SK_L}(s)} \over {q^3_{K_SK_L}(m_\phi)}} \cdot
 \Biggl| { {\Gamma_\phi m_\phi^3 \sqrt{\sigma(\phi\to K_SK_L)m_\phi}} \over
 {D_{\phi}(s)} } e^{i\phi_{K\overline{K}}}
\nonumber \\
 - { {\sqrt{\Gamma_\phi\Gamma_\omega m_\omega^3 m_\phi^2 6\pi
 B(\omega\to e^+e^-)B(\phi\to K_SK_L)}} \over {D_{\omega}(s)} } \\
 + { {\sqrt{\Gamma_\phi\Gamma_\rho m_\rho^3 m_\phi^2 6\pi
 B(\rho\to e^+e^-)B(\phi\to K_SK_L)}} \over {D_{\rho}(s)} }
 + A_{K_SK_L}^0  \Biggr|^2 ;
\nonumber
\end{eqnarray}
\begin{eqnarray}
\label{pi3sig}
 \sigma_{3\pi}(s) = { {1} \over {s^{3/2}} } {W(s) \over W(m_\phi)} \cdot
 \Biggl| 
 { {\Gamma_\phi m_\phi^2 \sqrt{\sigma(\phi\to 3\pi)m_\phi}} \over
 {D_{\phi}(s)} } e^{i\chi_{\phi\omega}} + \nonumber \\
 \sqrt[]{W(m_\phi) \over W(m_\omega)}
 { {\Gamma_\omega m_\omega^2 \sqrt{\sigma(\omega\to 3\pi)m_\omega}} \over
 {D_{\omega}(s)} } \sqrt[]{C_{\omega\rho\pi}(s)} +
 A^0_{3\pi} \Biggr|^2, 
\end{eqnarray}
 where
 $$
 \sigma(V\to X) = {{12\pi B(V\to e^+e^-)B(V\to X) } \over {m_V^2}},
 $$
 and $A^0_{K^+K^-}$, $A^0_{K_SK_L}$, $A^0_{3\pi}$ are complex constants,
 corresponding to the contributions of the  higher radial excitations of the
 $\rho$, $\omega$ and $\phi$ mesons in the cross section
 \cite{vysh1,vysh3,pdg}.

\section{Fitting of the experimental data}

 The experimental cross section $\sigma(s)$ 
 can be calculated as follows:
\begin{eqnarray}
\label{aprox}
 \sigma(s) = {{N(s)} \over {IL(s)\epsilon(s)(1+\delta_{rad}(s))}},
\end{eqnarray}
 where $N(s)$ is the number of selected events, $IL(s)$ is the 
 integrated luminosity, $\epsilon(s)$ is the detection efficiency,
 and $\delta_{rad}(s)$ is the radiative
 correction calculated according to  Ref.\cite{fadin}. The cross sections were
 fitted with the theoretical functions and the values of parameters and their
 errors were obtained using the procedure described in Ref.\cite{db} which
 takes into account the beam energy spread and the error in the
 beam energy determination.

 For the approximation of the $e^+e^- \to 3 \pi$ cross section, to
 take into account the detection efficiency dependence $\epsilon(E_\gamma)$ on
 the emitted $\gamma$-quantum  energy (Fig.\ref{pi3rad}), in the formula
 (\ref{aprox}) the factor $\epsilon(s)(1+\delta_{rad}(s))$ was replaced with
 the following expression:
\begin{eqnarray}
 {\int\limits^{E^{max}_\gamma}_0 \sigma(s,E_\gamma)F(s,E_\gamma)
 \epsilon(E_\gamma) \mathrm{d}E_\gamma \over {\sigma(s)}},
\end{eqnarray}
 where $E_\gamma$ is the emitted photon energy and $F(s,E_\gamma)$ is the
 electron ``radiator'' function \cite{fadin}.

\subsection{Fitting of the $e^+e^- \to \pi^+\pi^-\pi^0$ cross section}

 The fitting of the $e^+e^- \to \pi^+ \pi^- \pi^0$ cross section was performed
 with the $m_{\phi}$, $\Gamma_{\phi}$, $\sigma(\phi\to3\pi)$, $A^0_{3\pi}$ 
 and in some cases with the phase $\chi_{\phi-\omega}$ as free parameters.
 Approximations were performed under the following assumptions on  the
 $\chi_{\phi-\omega}$ value:
\begin{itemize}
\item
 $\chi_{\phi-\omega} = 180^\circ$,
\item
 $\chi_{\phi-\omega} = 180^\circ + \Delta \chi_{\phi-\omega}(s)$,
\item
 $\chi_{\phi-\omega}$ is a free parameter.
\end{itemize}
 Here the expression for $\Delta \chi_{\phi-\omega}(s)$ was taken from
 Ref.\cite{faza} and it is equal to
 $\Delta \chi_{\phi-\omega}(m_\phi) \simeq - 17^\circ$, when $s=m_\phi^2$.

\begin{figure}
\begin{center}
\epsfig{figure=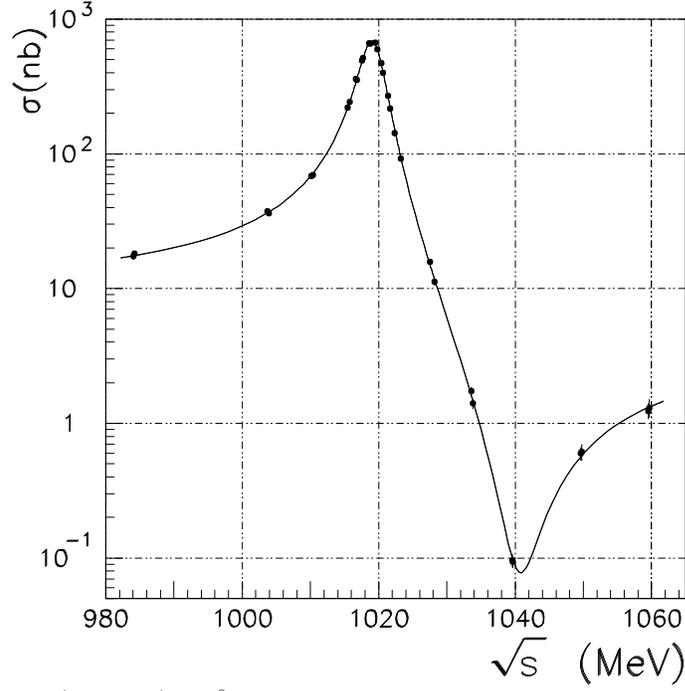,width=9cm}
\caption{The $e^+e^- \to \pi^+ \pi^- \pi^0$ cross section. The dots are
         experimental data, the curve is the fit.}
\label{pi3sig1}
\end{center}
\end{figure}

\begin{figure}
\begin{center}
\epsfig{figure=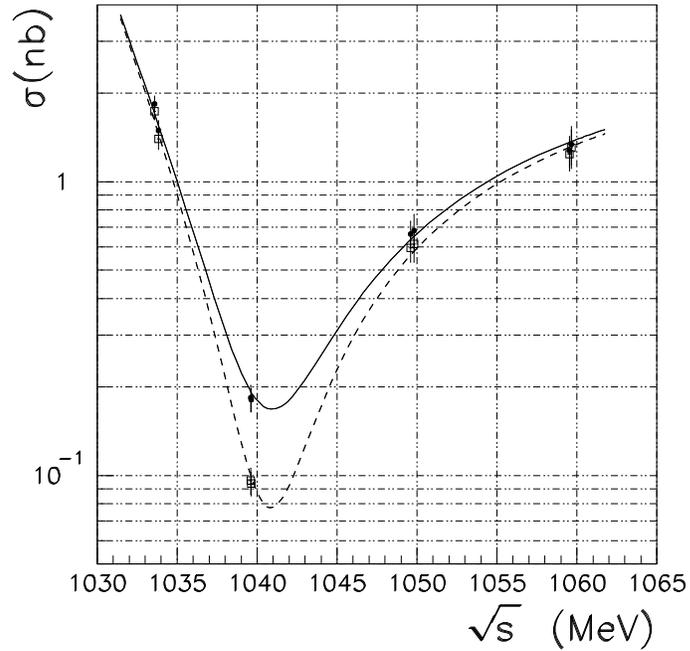,width=9cm}
\caption{The $e^+e^- \to \pi^+ \pi^- \pi^0$ cross section in the interference
         minimum region. Solid curve and dots: fit with
	 $\chi_{\phi-\omega}=180^\circ$, $\mbox{Im}(A^0_{3\pi})=0$; 
	 dashed curve and squares: the fit with 
	 $\chi_{\phi-\omega} = 180^\circ + \Delta \chi_{\phi-\omega}(s)$,
	 $\mbox{Im}(A^0_{3\pi})=0$.}
\label{pi3sig2}
\end{center}
\end{figure}

 The results of the fit are shown in Table~\ref{tab5}. The first three
 variants correspond to the case when both the real and imaginary parts of
 the  $A^0_{3\pi}$ amplitude were free parameters. The value of the
 $\mbox{Re}(A^0_{3\pi})$ differs from zero by more than three standard
 deviations. The value of $\mbox{Im}(A^0_{3\pi})$ is consistent with zero.
 Then the imaginary part was fixed at $\mbox{Im}(A^0_{3\pi})=0$ and only the
 $\mbox{Re}(A^0_{3\pi})$ was taken as a free parameter
 (four to six variants in Table~\ref{tab5}).
 The values of $m_{\phi}$ and $\Gamma_{\phi}$ do not depend on the
 theoretical model. The model dependence of the
 parameter $\sigma(\phi\to 3\pi)$
 is about $\sim 10\%$ and the difference of its values for variants 4 and 5
 is much greater than the errors. The energy dependence of the
 $e^+e^- \to 3\pi$
 cross section for the variant 5 is shown in Figure~\ref{pi3sig1}. At the right
 side from the resonance peak, the cross section has a minimum due to the
 interference of the  $\phi$--meson amplitude with the nonresonant part
 of the cross section.
 The cross section value does not actually depend on the model parameters in
 the energy range ~$\sqrt[]{s}= 984$ -- $1028$ MeV, but in the region of the
 interference minimum at ~$\sqrt[]{s}= 1040$ MeV, it can differ by a factor 
 of 2 for different models (Fig.\ref{pi3sig2}). This uncertainty is
 connected with the difference in the values of $\delta_{rad}(s)$ correction
 in Eq.(\ref{aprox}), calculated for different models.

\subsection{Fitting of the $e^+e^- \to K\overline{K}$ cross sections}

 The $e^+e^- \to K^+K^-$ and $K_SK_L$ cross
 sections were fitted with $m_\phi$, $\Gamma_\phi$,
 $\sigma(\phi\to K^+K^-)$, $\sigma(\phi\to K_SK_L)$ as free parameters.
 The parameters $A^0_{K^+K^-}$, $A^0_{K_SK_L}$ and $\phi_{K\overline{K}}$
 were either fixed ($\phi_{K\overline{K}}=180^\circ$ and
 $A^0_{K^+K^-}(A^0_{K_SK_L})=0$), or were free parameters. In particular,
 various fits were performed under the following assumptions:
\begin{enumerate}
\item
 $\phi_{K\overline{K}}=180^\circ$, ~~ $A^0_{K^+K^-}=A^0_{K_SK_L}=0$;
\item
 $\phi_{K\overline{K}}$ was free parameter,~~$A^0_{K^+K^-}=A^0_{K_SK_L}=0$;
\item
 $\phi_{K\overline{K}}$ and $\mbox{Re}(A^0_{K^+K^-})$, 
 $\mbox{Re}(A^0_{K_SK_L})$ were free parameters;
\item
 $\phi_{K\overline{K}}=180^\circ$, $\mbox{Re}(A^0_{K^+K^-})$,
 $\mbox{Re}(A^0_{K_SK_L})$, $\mbox{Im}(A^0_{K^+K^-})$, 
 $\mbox{Im}(A^0_{K_SK_L})$ were free parameters;
\item 
$\phi_{K\overline{K}}=180^\circ$,
$\mbox{Re}(A^0_{K^+K^-})$ and $\mbox{Re}(A^0_{K_SK_L})$ were free
      parameters.
\end{enumerate}
 The results of the fit are shown in Tables~\ref{tab41}~and~\ref{tab42}.

\begin{figure}
\begin{center}
\epsfig{figure=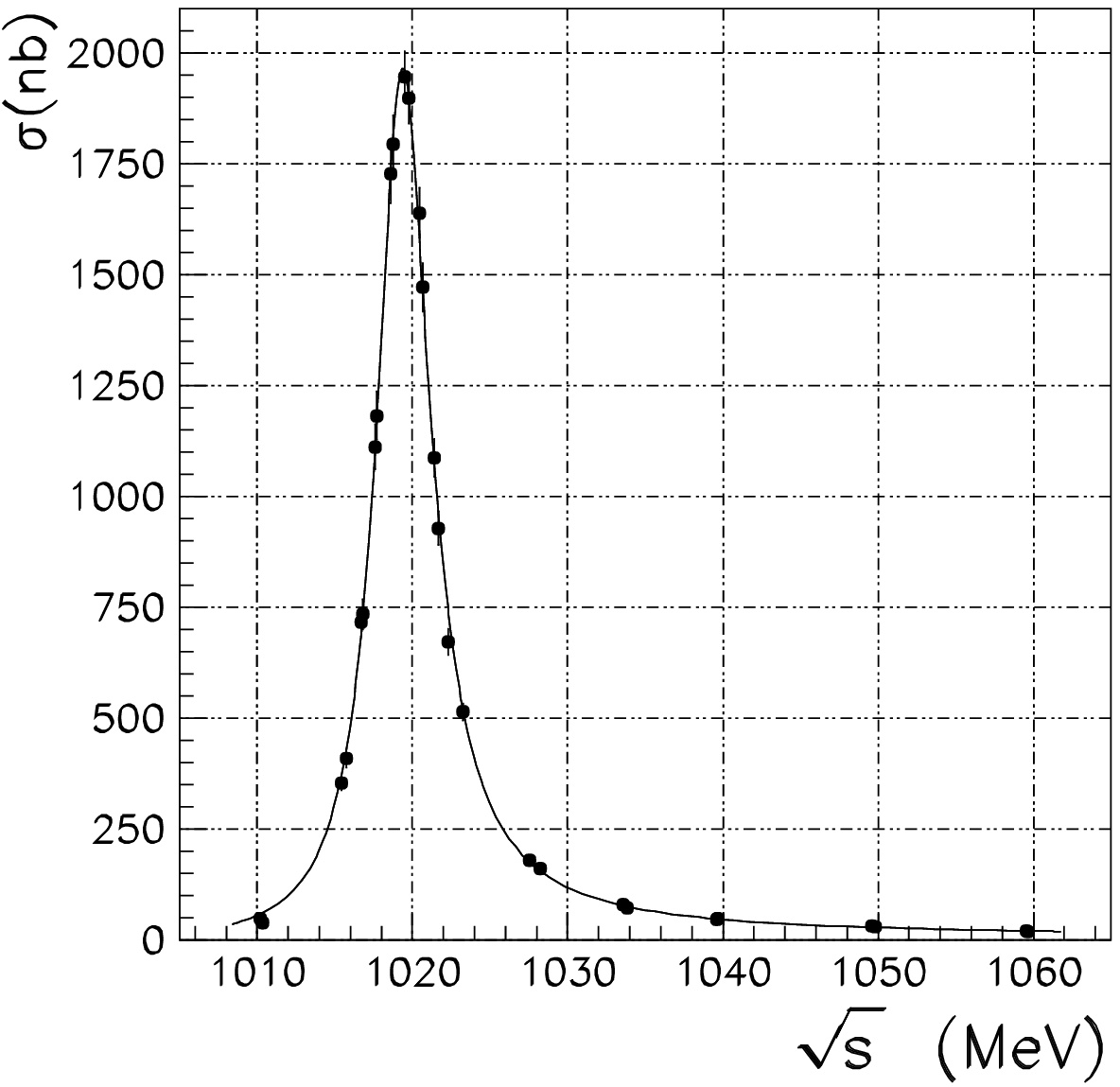,width=9cm}
\caption{The $e^+e^- \to K^+K^-$ cross section. The dots are experimental data,
         the curve is the fit.}
\label{kkcsig}
\end{center}
\end{figure}

 For the $e^+e^- \to K^+K^-$ process, the phase $\phi_{K\overline{K}}$
 agrees with the theoretically predicted value $180^\circ$. 
 The imaginary part of the $A^0_{K^+K^-}$ amplitude is consistent with zero,
 and its real part is different from zero by about only one standard deviation
 (variants 4 and 5 in Table~\ref{tab41}). The values of $m_{\phi}$ and
 $\Gamma_{\phi}$ do not depend on the applied theoretical model. The value of
 $\sigma(\phi\to K^+K^-)$ varies by 2\% for different variants, but these
 biases are less than its error. The cross section is almost independent of
 the applied model (Fig.\ref{kkcsig}).

\begin{figure}
\begin{center}
\epsfig{figure=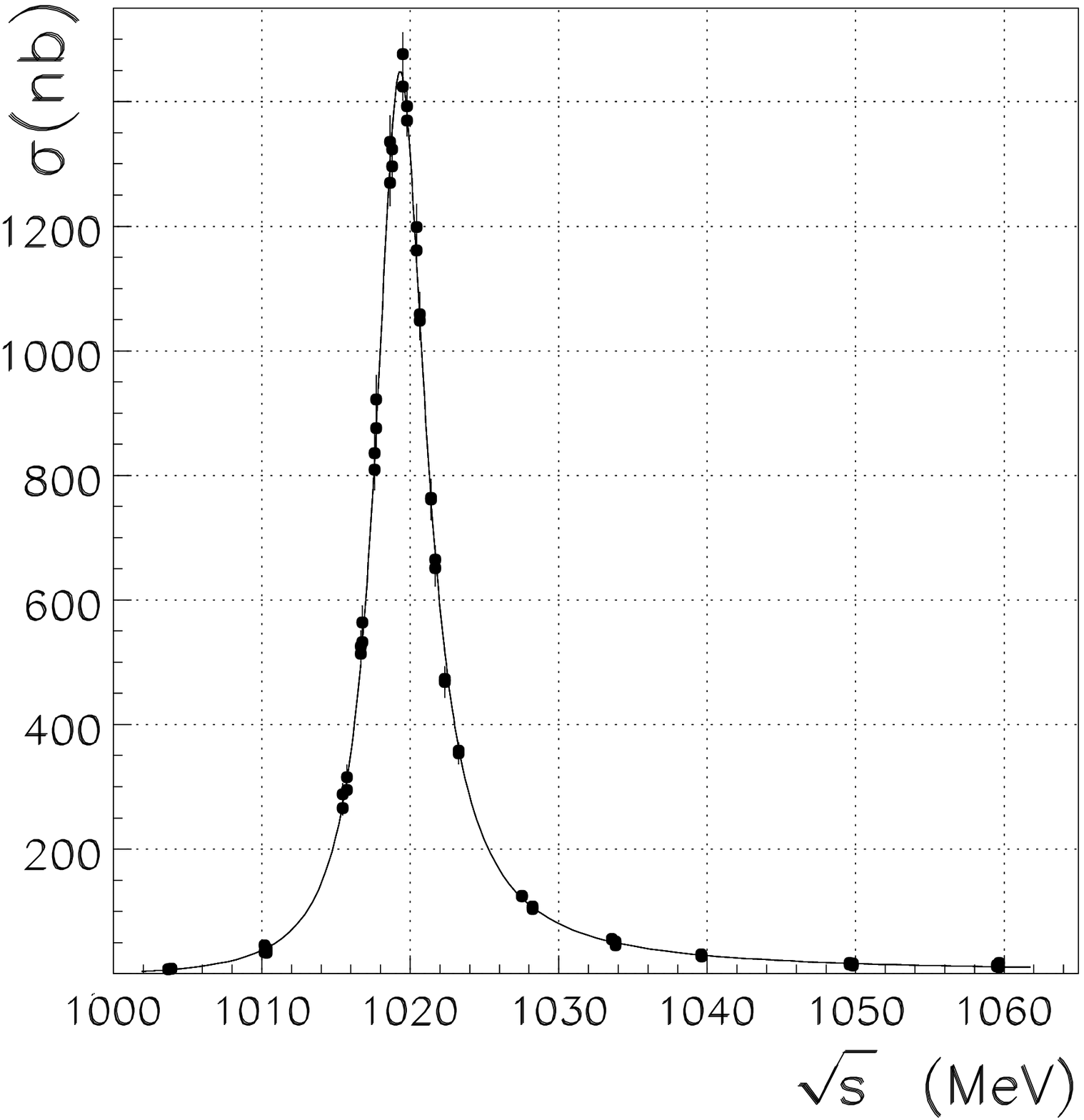,width=9cm}
\caption{The $e^+e^- \to K_SK_L$ cross section. The dots are experimental data,
         the curve is the fit.}
\label{kslsig}
\end{center}
\end{figure}

 The data for the $e^+e^- \to K_SK_L$ reaction obtained in charged and neutral
 modes were fitted simultaneously. For this process 
 the phase $\phi_{K\overline{K}}$ is
 different from the expected $180^\circ$ by more then five standard deviations
 and is close to $90^\circ$ (variant 2 in Table~\ref{tab42}).
 The imaginary part of the $A^0_{K_SK_L}$ amplitude
 is consistent with zero, and its real part is different from zero 
 by about two standard deviations (variants 5 and 4 in Table~\ref{tab42}).
 The parameters $m_{\phi}$ and $\Gamma_{\phi}$ are slightly dependent on the
 theoretical assumptions. The value of $\sigma(\phi\to K_SK_L)$ varies by 2\%
 for different variants and is shown in Fig.\ref{kslsig}. The cross section is
 almost insensitive to the choice of the model.

\subsection{Combined fitting of the cross sections}

 To obtain the $\phi$-meson production cross section and branching 
 ratios of the main decay modes, the combined fitting of the experimental
 data, taking into account the $e^+e^- \to \eta\gamma$ cross section also, was
 carried out. As this work does not contain the analysis of the
 $e^+e^- \to \eta \gamma$ reaction, it was natural to use the cross section
 $\sigma(\phi \to \eta \gamma)$ from the previous  SND measurements presented
 in  Ref.\cite{bad1}. In the Ref. \cite{bad1} the correction
 $\delta_{over}=0.97 \pm 0.03$ to the detection efficiency was not applied.
 The value  $\sigma(\phi \to \eta \gamma) = 53.2 \pm 2$ nb, presented
 there, was increased
 by 3\% and the value $54.8 \pm 2.1$ nb was used in the fit.
  The results of the fit,
 for the different model assumptions, are shown in Table~\ref{tab6}. 
 In all fits $\mbox{Im}(A^0_{K^+K^-}) = 0$, $\mbox{Im}(A^0_{K_SK_L}) = 0$,
 and $\mbox{Im}(A^0_{3\pi}) = 0$ were set.

 Using these data, the following $\phi$-meson parameters were obtained:
 the $\phi$-meson production cross section
\begin{eqnarray}
\label{bee1}
 \sigma(\phi) = \sum_X{\sigma(\phi \to X)}, X=K^+K^-, K_SK_L, \pi^+\pi^-\pi^0,
 \eta \gamma
\end{eqnarray}
 the branching ratios of the main decay modes
\begin{eqnarray}
\label{bee2}
 B(\phi \to X) = {{\sigma(\phi \to X)} \over {\sigma(\phi)}},
\end{eqnarray}
 and the branching ratio of the decay into an $e^+e^-$ pair:
\begin{eqnarray}
\label{bee3}
 B(\phi \to e^+e^-) = {{\sigma(\phi) m_\phi^2} \over {12 \pi}}.
\end{eqnarray}
 
The final results are
\begin{eqnarray}
 B(\phi \to e^+e^-) = (2.93 \pm 0.02 \pm 0.14 \pm 0.02) \times 10^{-4},
\nonumber
\end{eqnarray}
\begin{eqnarray}
 B(\phi \to K^+K^-) = 47.6 \pm 0.3 \pm 1.6 \pm 0.3 \%
\nonumber
\end{eqnarray}
\begin{eqnarray}
 B(\phi \to K_SK_L) = 35.1 \pm 0.2 \pm 1.2 \pm 0.3 \%
\nonumber
\end{eqnarray}
\begin{eqnarray}
 B(\phi \to 3\pi) = 15.9 \pm 0.2 \pm 0.7 \pm 0.4 \%
\nonumber
\end{eqnarray}
\begin{eqnarray}
 B(\phi \to \eta \gamma) = 1.33 \pm 0.03 \pm 0.05 \pm 0.01 \%
\nonumber
\end{eqnarray}
 These values correspond to the the fit where 
 $\chi_{\phi-\omega} = 180^\circ + \Delta\chi_{\phi-\omega}(s)$,
 $\phi_{K\overline{K}}=180^\circ$, $\mbox{Im}(A^0_{3\pi})=0$, $A_{K_SK_L}^0=0$
 and $A_{K^+K^-}^0 = 0$ (variant 2 in Table~\ref{tab6}). The third
 error is  a model uncertainty. The presented values are in good agreement
 with the world average and with other experimental data (Table~\ref{tab10}).
 Using the SND data, the following coupling constants were calculated:
 $|g_{\phi\gamma}| = 6538 \pm 156$ MeV$^2$,~
 $|g_{\phi K^+K^-}| = 4.391 \pm 0.078$,~ 
 $|g_{\phi K_SK_L}| = 4.664 \pm 0.086$,~
 $|g_{\phi\rho\pi}| = 0.815 \pm 0.021$ GeV$^{-1}$, and 
 $|g_{\phi\eta\gamma}| = 0.0593 \pm 0.0013$ GeV$^{-1}$.

 The measured cross sections are given in 
 Table~\ref{tab7}. The systematic errors of the $e^+e^- \to K^+K^-$,
 $e^+e^- \to \pi^+\pi^-\pi^0$ cross sections and of the related parameters 
 $\sigma(\phi\to K^+K^-)$, $\sigma(\phi \to 3\pi)$ are 7.1\% and 5\%
 respectively and include the systematic uncertainties in the 
 detection efficiency
 and luminosity determinations. For the $e^+e^- \to K_SK_L$ cross section, the 
 systematic error is 4.2\% in the charged mode and 4.0\% in the neutral mode.
 The systematic error of the parameter $\sigma(\phi\to K_SK_L)$ is
 3.2 \% .

\section{Discussion}

 The $\phi$-meson mass and width measured in the three studied main decay
 modes are consistent with each other. The measurements of the $\phi$-meson
 parameters in all channels have high accuracy. Let us mention that all
 these processes have different energy dependence for detection efficiency.
 The dependence of the mass value on the
 applied theoretical model is less than the measurement accuracy. As a result
 of the combined approximation, we find:
 $$m_{\phi} = 1019.42 \pm 0.02 \pm 0.04 \mbox{~MeV},$$
 where the systematic error of 0.04 MeV is attributed to the possible common
 shift of the collider energy scale. This value is consistent with the world
 average: $1019.417 \pm 0.014$ MeV \cite{pdg}.
 The value of the total $\phi$-meson width, obtained by using different
 models, can vary from  4.18 to 4.21 MeV and was found to be
 $$\Gamma_{\phi} = 4.21 \pm 0.03 \pm 0.02 \mbox{~MeV}$$
 (the systematic error 0.02 MeV is due to model dependence), and differs from
 the world average $4.458 \pm 0.032$ MeV \cite{pdg} by 4.8 standard deviations
 and from the most precise measurement
 $\Gamma_{\phi} = 4.477 \pm 0.036 \pm 0.022$ MeV \cite{cmd3} by 4.5
 standard deviations. 

 The $e^+e^- \to \pi^+\pi^-\pi^0$ process, as the approximation showed,
 cannot be described by only $\phi$ and $\omega$ meson decays and it is
 necessary to include additional amplitude $A^0_{3\pi}$, which can be explained
 as a contribution of the higher resonances. The presence of such an amplitude
 is also expected from 
 the experimental data on the $e^+e^- \to \pi^+\pi^-\pi^0$ cross
 section measured by the SND detector \cite{sis} in the energy region above the
 $\phi$-resonance, which cannot be described with the conventional VDM.
 The value of the amplitude $A^0_{3\pi}=0.046 \pm 0.003$~MeV$^{1/2}$
 (Table~\ref{tab6}) is only 2 times lower than the real part of
 $\omega$-meson amplitude in the vicinity of the $\phi$-meson peak
 $\mbox{Re}(A_\omega) \simeq 0.09$~MeV$^{1/2}$ and approximately equals its
 imaginary part $\mbox{Im}(A_\omega) \simeq 0.05$~MeV$^{1/2}$.
 The presence of the $A^0_{3\pi}$ amplitude prevents a test of the deviation
 of the $\chi_{\omega-\phi}$ phase from $180^\circ$. Both 
 $\chi_{\phi-\omega} = 180^\circ$ and
 $180^\circ + \Delta\chi_{\phi-\omega}(s)$ \cite{faza} are consistent with
 the experimental data. If the $\chi_{\phi-\omega}$ phase is considered to be
 a free parameter with no energy dependence, then its fitted value is
 $178^\circ \pm 27^\circ$ for free real and imaginary parts of $A^0_{3\pi}$  
 (variant 5 in Table~\ref{tab5}), and $165^\circ {\pm^{19}_{7}}^\circ$, if 
 $\mbox{Im}(A^0_{3\pi})=0$ is set (variant 3 in Table~\ref{tab6}).

 The parameter :
 $$\sigma(\phi \to 3\pi) = 659 \pm 6 \pm 33 \pm^{20}_{5} \mbox{~nb}$$
 has a rather large model error ($\pm^{20}_5$ nb ), due to the uncertainty 
 in the choice of the
 phase $\chi_{\omega-\phi}$. The quoted value of the  
 $\sigma(\phi \to 3\pi)$ corresponds to the choice
 $\chi_{\omega-\phi} = 180^\circ + \Delta\chi_{\phi-\omega}(s)$ and
 $\mbox{Im}(A^0_{3\pi})=0$. It agrees with other measurements,
 for example, $\sigma(\phi \to 3\pi) = 619 \pm 39 \pm 12$ from Ref.\cite{cmd2}
 and $\sigma(\phi \to 3\pi) = 654 \pm 26 \pm 30$ from Ref.\cite{cmd1}.

 In case of the $e^+e^- \to K^+K^-$ and $K_SK_L$ processes, the
 $\mbox{Im}(A_{K^+K^-}^0)$, $\mbox{Im}(A_{K_SK_L}^0)$ agree with zero
 (variant 4 in Tables~\ref{tab41} and \ref{tab42}) and 
 $\mbox{Re}(A_{K^+K^-}^0)$, $\mbox{Re}(A_{K_SK_L}^0)$ deviates from zero
 by about two standard deviations (variant 5, Table~\ref{tab6}).
 The phase $\phi_{K\overline{K}}$ extracted from the combined fit,
 or from the  fitting of the $e^+e^- \to K^+K^-$ process, is consistent with
 $180^\circ$
 (variants 2 and 4 in Tables~\ref{tab41} and \ref{tab6}),
 but the fitting of the $e^+e^- \to K_SK_L$ data gives the phase value close to
 $90^\circ$. Such a result can be attributed to the uncertainty of the 
 nonresonant
 contributions to the cross section  and was used to estimate the
 model error in the cross section determination. The results 
 $$\sigma(\phi\to K^+K^-) = 1968 \pm 20 \pm 140$$
 $$\sigma(\phi\to K_SK_L) = 1451 \pm 10 \pm 48 \mbox{~nb}$$ agree with the
 world average \cite{pdg}.

 Isotopic symmetry predicts the equality of the
 $g_{\phi K^+K^-}$ and $g_{\phi K_SK_L}$
 coupling constants. The electromagnetic interactions of
 charged kaons in the final state lead to the relation
 $g_{\phi K^+K^-}/\sqrt[]{Z(m_\phi)}=g_{\phi K_SK_L}$. Using the Particle 
 Data Group (PDG) data,
 \cite{pdg} one can find:
\begin{eqnarray}
\label{gg11}
 {g_{\phi K^+K^-} \over g_{\phi K_SK_L}} {1 \over \sqrt[]{Z(m_\phi)}} =
 \sqrt[]{\Biggl({{B(\phi \to K^+K^-) q_{K_SK_L}^3(m_\phi)} \over
  {B(\phi \to K_SK_L) q_{K^+K^-}^3(m_\phi)}} \Biggr)
  {1 \over {Z(m_\phi)}}} = 0.95 \pm  0.01
\end{eqnarray}
 This value differs considerably (five standard deviations) from the expected
 value of unity. It is possible that the Coulomb correction is less than
 expected due to finite dimensions of the final particles. A similar problem
 in the $\Upsilon_{4s} \to B\overline{B}$ decay was discussed in
 Refs.\cite{bbform1,bbform2}. Quite recently the ratio of the constants 
 $g_{\phi K^+K^-}$ and $g_{\phi K_SK_L}$ was discussed in Ref.\cite{bramon},
 where it was suggested to study this ratio in $e^+e^-$ annihilation around
 the $\phi$ resonance peak. The drawback of this proposal is a very   
 low sensitivity of such studies to the nonresonant contribution to the total
 cross section. The extracted coupling constant values depend on
 the nonresonant contribution to the total cross section from the
 $\rho$ and $\omega$ mesons and from higher radial excitations
 ($\rho^\prime$, $\omega^\prime$, $\phi^\prime$, ...). So one of the
 possible approaches to the solution of this problem can be a precise study
 of these contributions to the total cross section, by means of the measurement
 of the $e^+e^- \to K\overline{K}$ cross section in a wide energy range.
 Unfortunately the SND data have rather high systematic errors, which preclude
 such studies, and they are accumulated in
 a rather narrow energy region. From SND data (Table~\ref{tab6}) one can obtain
\begin{eqnarray}
 {g_{\phi K^+K^-} \over g_{\phi K_SK_L}} {1 \over \sqrt[]{Z(m_\phi)}} =
 0.92 \pm 0.03.
\end{eqnarray}
 This value agrees with the quoted world average, but is also consistent
 with the expected value of 1 ( the difference is about 2.5 standard
 deviations).

\section{Conclusion}

 The main parameters of the $\phi$-resonance were obtained in experiments
 with the SND detector at the VEPP-2M collider. The integrated luminosity used
 was about $8.5~\mbox{pb}^{-1}$. The processes $e^+e^- \to K^+K^-$,
 ~ $e^+e^- \to K_SK_L$ and $e^+e^- \to \pi^+\pi^-\pi^0$ were studied.
 The measured cross sections were approximated within the VDM, taking
 into account $\rho$, $\omega$ and $\phi$-mesons. Possible contributions
 from higher resonances $\rho^\prime$, $\omega^\prime$, $\phi^\prime$ were
 included as constant amplitudes. The $\phi$-meson parameters obtained by the
 approximation  (Table~\ref{tab10}) mainly agree with PDG data and have
 accuracies comparable with the world averages.
 The  $\phi$-meson total width was found to be $4.21 \pm 0.04$ MeV in
 contradiction to the PDG value $4.458 \pm 0.032$ MeV. For a good
 approximation of the $e^+e^- \to \pi^+\pi^-\pi^0$ cross section the
 additional contribution is strongly required, and its value is about 2
 times lower than the real part of the $\omega$-meson contribution. This
 contribution can be attributed to the higher resonances, for example to the
 resonance structure found by SND \cite{sis}.

 The authors are grateful to N.N.Achasov, E.P.Solodov and S.I.Eidelman for
 useful discussions.

\newpage

\begin{table}[h]
\begin{center}
\begin{tabular}[t]{cccccccc}
\multicolumn{4}{c}{Scan PHI9801} &
\multicolumn{4}{c}{Scan PHI9802} \\
 $\sqrt[]{s}$&$\sigma$&$IL_{cha}$&$IL_{neu}$&
 $\sqrt[]{s}$& $\sigma$&$IL_{cha}$&$IL_{neu}$  \\
(MeV)&(MeV)&$(\mbox{nb}^{-1})$&$(\mbox{nb}^{-1})$&
(MeV)&(MeV)&$(\mbox{nb}^{-1})$&$(\mbox{nb}^{-1})$ \\ \hline
 984.21$\pm$0.10&0.36&173.7$\pm$2.3&159.7$\pm$2.4&
 984.02$\pm$0.10&0.36&200.9$\pm$2.7&186.7$\pm$2.7\\
1003.91$\pm$0.10&0.36&209.9$\pm$2.8&193.6$\pm$2.8&
1003.71$\pm$0.10&0.36&181.1$\pm$2.4&171.4$\pm$2.6\\
1010.17$\pm$0.13&0.37&156.0$\pm$2.1&140.8$\pm$2.2&
1010.34$\pm$0.13&0.37&158.1$\pm$2.1&147.6$\pm$2.2\\
1015.75$\pm$0.07&0.37&168.2$\pm$2.2&160.1$\pm$2.4&
1015.43$\pm$0.07&0.39&200.9$\pm$2.7&183.7$\pm$2.7\\
1016.68$\pm$0.07&0.37&304.7$\pm$4.0&286.8$\pm$4.1&
1016.78$\pm$0.07&0.38&333.0$\pm$4.4&299.1$\pm$4.2\\
1017.59$\pm$0.07&0.37&483.3$\pm$6.3&444.2$\pm$6.1&
1017.72$\pm$0.07&0.37&528.4$\pm$6.9&478.7$\pm$6.6\\
1018.78$\pm$0.07&0.38&527.0$\pm$6.9&482.8$\pm$6.6&
1018.62$\pm$0.07&0.38&536.5$\pm$7.0&484.1$\pm$6.6\\
1019.79$\pm$0.07&0.37&560.2$\pm$7.3&511.4$\pm$7.0&
1019.51$\pm$0.07&0.37&523.6$\pm$6.9&472.7$\pm$6.5\\
1020.65$\pm$0.08&0.37&323.7$\pm$4.3&296.3$\pm$4.2&
1020.43$\pm$0.08&0.37&355.6$\pm$4.7&330.7$\pm$4.6\\
1021.68$\pm$0.08&0.37&163.2$\pm$2.2&150.4$\pm$2.3&
1021.41$\pm$0.08&0.37&184.3$\pm$2.4&172.1$\pm$2.6\\
1023.27$\pm$0.09&0.37&202.6$\pm$2.7&186.2$\pm$2.8&
1022.32$\pm$0.09&0.39&178.1$\pm$2.4&166.3$\pm$2.5\\
1028.23$\pm$0.14&0.37&167.9$\pm$2.2&150.1$\pm$2.3&
1027.52$\pm$0.14&0.38&218.4$\pm$2.9&211.5$\pm$3.1\\
1033.84$\pm$0.10&0.37&167.8$\pm$2.2&147.7$\pm$2.3&
1033.58$\pm$0.10&0.36&182.2$\pm$2.4&172.3$\pm$2.6\\
1039.59$\pm$0.10&0.39&162.7$\pm$2.1&151.5$\pm$2.3&
1039.64$\pm$0.10&0.37&185.1$\pm$2.5&169.2$\pm$2.5\\
1049.81$\pm$0.10&0.38&196.9$\pm$2.6&178.9$\pm$2.7&
1049.60$\pm$0.10&0.38&187.2$\pm$2.5&176.3$\pm$2.6\\
1059.66$\pm$0.10&0.38&169.8$\pm$2.3&158.6$\pm$2.4&
1059.52$\pm$0.10&0.50&216.3$\pm$2.9&205.6$\pm$3.0\\
\end{tabular}
\caption{Experimental main parameters: $\sqrt[]{s}$ -- the energy in the center
         of mass system, $\sigma$ -- beam energy spread,
	 $IL_{cha}$, $IL_{neu}$ -- integrated luminosities used for normalizing
	 the events containing charged particles and with neutral particles 
only.}
\label{tab1}
\end{center}
\end{table}

\begin{table}[t]
\begin{center}
\begin{tabular}[t]{cccccc}
\multicolumn{3}{c}{Scan PHI9801} &
\multicolumn{3}{c}{Scan PHI9802} \\ 
 $\sqrt[]{s}$ (MeV) & $N_{K^+K^-}$ & $\epsilon_{K^+K^-}$ &
 $\sqrt[]{s}$ (MeV) & $N_{K^+K^-}$ & $\epsilon_{K^+K^-}$   \\ \hline
 1010.17&306$\pm$23&0.055$\pm$0.012&1010.34&272$\pm$22&0.0599$\pm$0.012 \\
 1015.75&8370$\pm$108&0.1669$\pm$0.0045&1015.43&8524$\pm$110&0.1638$\pm$
 0.0045 \\
 1016.68&27723$\pm$195&0.1749$\pm$0.0046&1016.78&31181$\pm$207&0.1758$\pm$
 0.0046 \\
 1017.59&69921$\pm$321&0.1814$\pm$0.0047&1017.72&72263$\pm$325&0.1821$\pm$
 0.0047 \\
 1018.78&124598$\pm$431&0.1878$\pm$0.0049&1018.62&121718$\pm$421&0.1869$\pm$
 0.0048 \\
 1019.79&149331$\pm$472&0.1922$\pm$0.0049&1019.51&139295$\pm$452&0.1909$\pm$
 0.0049\\
 1020.65&73994$\pm$321&0.1958$\pm$0.005&1020.43&88199$\pm$342&0.1948$\pm$
 0.005 \\
 1021.68&26342$\pm$187&0.2001$\pm$0.0052&1021.41&33916$\pm$211&0.1992$\pm$
 0.0051 \\
 1023.27&20637$\pm$167&0.2037$\pm$0.0053&1022.32&22111$\pm$169&0.2021$\pm$
 0.0052 \\
 1028.23&7009$\pm$97&0.2115$\pm$0.0059&1027.52&9865$\pm$112&0.2106$\pm$
 0.0059 \\
 1033.84&3747$\pm$72&0.2163$\pm$0.0064&1033.58&4437$\pm$75&0.2164$\pm$
 0.0064 \\
 1039.59&2349$\pm$52&0.1975$\pm$0.0063&1039.64&2745$\pm$58&0.1972$\pm$
 0.0063 \\
 1049.81&1270$\pm$40&0.1330$\pm$0.005&1049.6&1298$\pm$40&0.1334$\pm$
 0.005 \\
 1059.66&672$\pm$27&0.1228$\pm$0.0053&1059.52&961$\pm$33&0.1233$\pm$
 0.0053 \\
\end{tabular}
\caption{Event numbers $N$ and detection efficiencies $\epsilon$ for
       the $e^+e^- \to K^+K^-$ process versus energy.}
\label{tab2}
\end{center}
\end{table}

\begin{table}[h]
\begin{center}
\begin{tabular}[t]{cccccc}
\multicolumn{3}{c}{Scan PHI9801} &
\multicolumn{3}{c}{Scan PHI9802} \\ 
 $\sqrt[]{s}$ (MeV) & $N_{K_SK_L}$ & $\epsilon_{K_SK_L}$ &
 $\sqrt[]{s}$ (MeV) & $N_{K_SK_L}$ & $\epsilon_{K_SK_L}$   \\ \hline
 1003.91&56$\pm$9&0.0475$\pm$0.019&1003.71&46$\pm$9&0.0469$\pm$0.019 \\ 
 1010.17&277$\pm$18&0.532$\pm$0.0069&1010.34&249$\pm$17&0.0534$\pm$0.0069 \\ 
 1015.75&2372$\pm$50&0.0614$\pm$0.0028&1015.43&2594$\pm$53&0.0614$\pm$
 0.0028 \\ 
 1016.68&7092$\pm$88&0.0612$\pm$0.0016&1016.78&8314$\pm$95&0.0612$\pm$
 0.0016 \\ 
 1017.59&18335$\pm$141&0.0634$\pm$0.0011&1017.72&19545$\pm$145&0.0632$\pm$
 0.0011 \\ 
 1018.78&30820$\pm$182&0.0631$\pm$0.0011&1018.62&31703$\pm$185&0.0631$\pm$
 0.0011 \\ 
 1019.79&36358$\pm$198&0.0636$\pm$0.0012&1019.51&35272$\pm$195&0.0636$\pm$
 0.0012 \\ 
 1020.65&17276$\pm$137&0.0632$\pm$0.0011&1020.43&21071$\pm$150&0.0633$\pm$
 0.0011 \\ 
 1021.68&5975$\pm$80&0.0642$\pm$0.0019&1021.41&7726$\pm$90&0.0643$\pm$
 0.0019 \\ 
 1023.27&4453$\pm$69&0.0633$\pm$0.0022&1022.32&4895$\pm$72&0.0637$\pm$
 0.0022 \\ 
 1028.23&1361$\pm$39&0.0627$\pm$0.0035&1027.52&2079$\pm$47&0.0630$\pm$
 0.0035 \\ 
 1033.84&679$\pm$27&0.0607$\pm$0.0043&1033.58&901$\pm$31&0.0608$\pm$0.0043 \\ 
 1039.59&471$\pm$23&0.0620$\pm$0.0057&1039.64&517$\pm$24&0.062$\pm$0.0056 \\
 1049.81&311$\pm$19&0.0633$\pm$0.0074&1049.60&330$\pm$20&0.062$\pm$0.0073 \\
 1059.66&234$\pm$17&0.0395$\pm$0.0069&1059.52&245$\pm$17&0.0395$\pm$0.0069 \\
\end{tabular}
\caption{Event numbers $N$ and detection efficiencies $\epsilon$ for the
         $e^+e^- \to K_SK_L$ process with $K_S\to\pi^+\pi^-$ decay.}
\label{tab21}
\end{center}
\end{table}

\begin{table}[h]
\begin{center}
\begin{tabular}[t]{cccc}
\multicolumn{2}{c}{Scan PHI9801} &
\multicolumn{2}{c}{Scan PHI9802} \\ 
 $\sqrt[]{s}$ (MeV) & $N_{K_SK_L}$ & $\sqrt[]{s}$ (MeV)&$N_{K_SK_L}$ \\ \hline
 1003.91&   74$\pm$9  &1003.71&   57$\pm$  8 \\ 
 1010.17&  241$\pm$16 &1010.34&  250$\pm$ 16 \\ 
 1015.75& 2340$\pm$50 &1015.43& 2431$\pm$ 51 \\ 
 1016.68& 7267$\pm$87 &1016.78& 7865$\pm$ 90 \\ 
 1017.59&17558$\pm$135&1017.72&20422$\pm$146 \\ 
 1018.78&29870$\pm$176&1018.62&29376$\pm$175 \\ 
 1019.79&35002$\pm$190&1019.51&32910$\pm$185 \\ 
 1020.65&16908$\pm$132&1020.43&20454$\pm$145 \\ 
 1021.68& 5968$\pm$79 &1021.41& 7680$\pm$ 89 \\ 
 1023.27& 4463$\pm$68 &1022.32& 4945$\pm$ 72 \\ 
 1028.23& 1376$\pm$38 &1027.52& 2187$\pm$ 48 \\ 
 1033.84&  767$\pm$28 &1033.58&  936$\pm$ 31 \\ 
 1039.59&  521$\pm$24 &1039.64&  559$\pm$ 24 \\
 1049.81&  372$\pm$20 &1049.60&  390$\pm$ 20 \\
 1059.66&  234$\pm$16 &1059.52&  333$\pm$ 19 \\
\end{tabular}
\caption{Event numbers $N$ for the $e^+e^- \to K_SK_L$ process with
 $K_S\to\pi^0\pi^0$ decay.}
\label{tab22}
\end{center}
\end{table}

\begin{table}[h]
\begin{center}
\begin{tabular}[t]{cccccc}
\multicolumn{3}{c}{Scan PHI9801} &
\multicolumn{3}{c}{Scan PHI9802} \\
 $\sqrt[]{s}$ (MeV) & $N_{3\pi}$ &$N_{bkg}$ &
 $\sqrt[]{s}$ (MeV) & $N_{3\pi}$ &$N_{bkg}$  \\ \hline
  984.21 & 498   $\pm$ 23  &5   &  984.02 &   552 $\pm$  25 &   5 \\
 1003.91 & 1145  $\pm$ 36  &8   & 1003.71 &  1026 $\pm$  34 &   7 \\
 1010.17 & 1548  $\pm$ 42  &11  & 1010.34 &  1592 $\pm$  43 &  12 \\
 1015.75 & 5610  $\pm$ 79  &52  & 1015.43 &  6094 $\pm$  83 &  55 \\
 1016.68 & 14833 $\pm$ 129 &154 & 1016.78 & 15956 $\pm$ 133 & 177 \\
 1017.59 & 31862 $\pm$ 190 &394 & 1017.72 & 36261 $\pm$ 202 & 458 \\
 1018.78 & 45502 $\pm$ 228 &657 & 1018.62 & 46722 $\pm$ 230 & 643 \\
 1019.79 & 46590 $\pm$ 229 &770 & 1019.51 & 47414 $\pm$ 232 & 735 \\
 1020.65 & 20172 $\pm$ 151 &381 & 1020.43 & 25387 $\pm$ 169 & 448 \\
 1021.68 & 6339  $\pm$ 86  &144 & 1021.41 &  8588 $\pm$  99 & 180 \\
 1023.27 & 4160  $\pm$ 69  &104 & 1022.32 &  4969 $\pm$  76 & 121 \\
 1028.23 & 444   $\pm$ 22  &4   & 1027.52 &   715 $\pm$  27 &   5 \\
 1033.84 & 203   $\pm$ 16  &3   & 1033.58 &   254 $\pm$  16 &   4 \\
 1039.59 & 118   $\pm$ 12  &2   & 1039.64 &   133 $\pm$  12 &   2 \\
 1049.81 & 91    $\pm$ 12  &2   & 1049.60 &    87 $\pm$   9 &   2 \\
 1059.66 & 54    $\pm$ 8   &4   & 1059.52 &    66 $\pm$   8 &   5 \\
\end{tabular}
\caption{Event numbers $N_{3\pi}$ and $N_{bkg}$ of the 
$e^+e^- \to \pi^+\pi^-\pi^0$ (after background subtraction)
and background processes respectively. The accuracy of background event
numbers determination is less than 15\% at all energy points.}
\label{tab3}
\end{center}
\end{table}

\begin{table}[h]
\begin{center}
\begin{tabular}[t]{ccccccc}
 $N$&1&2&3&4&5&6 \\
 $m_{\phi}-1000$&$19.41 \pm 0.03$&$19.42 \pm 0.03$&$19.41 \pm 0.03$&
            $19.41 \pm 0.03$&$19.42 \pm 0.03$&$19.42 \pm 0.03$ \\
 (MeV)& & & & & & \\ 
 $\Gamma_{\phi}$&$4.18 \pm 0.06$&$4.18 \pm 0.06$&$4.18  \pm 0.06$&
                 $4.18 \pm 0.06$&$4.19 \pm 0.06$&$4.18  \pm 0.06$ \\
 (MeV)& & & & & & \\ 
 $\sigma(\phi\to 3\pi)$&$668 \pm^{47}_{16}$&$657 \pm^{33}_{16}$&$668 \pm 24$&
                        $704 \pm 8$&$661 \pm 7$&$666 \pm^{52}_{18}$ \\
 (nb)& & & & & & \\ 
 $\mbox{Re}(A^0_{3\pi}) \cdot 10^3$&$-54 \pm^{10}_{5}$&$-48 \pm^{15}_{6}$&
 $-53 \pm 6$&$-47 \pm 3$&$-47 \pm 3$&$-47 \pm 3$ \\
 (MeV$^{1/2}$)& & & & & & \\ 
 $\mbox{Im}(A^0_{3\pi}) \cdot 10^3$&$-35 \pm^{40}_{14}$&$-4 \pm^{30}_{10}$&
 $-33 \pm 60$&$0$&$0$&$0$ \\
 (MeV$^{1/2}$)& & & & & & \\ 
 $\chi_{\phi-\omega}$&180&$\Psi_{\phi-\omega}(s)$&$178 \pm 27$&180&
 $\Psi_{\phi-\omega}(s)$&$165 \pm^{20}_{6}$ \\
 (degree)& && & &&
 \\ 
 $\chi^2/N_{df}$&20.39/27&20.46/27&20.39/26&20.56/28&20.49/28&20.49/27
 \\ \hline
\end{tabular}
\normalsize
\caption{Fit results for the process $e^+e^- \to \pi^+\pi^-\pi^0$.
         The column number $N$ corresponds to the different variants of
         choice of the
         phase $\chi_{\phi-\omega}$ and
         the imaginary part of the amplitude
	 $\mbox{Im}(A^0_{3\pi})$. Here
	 $\Psi_{\phi-\omega}(s) = 180^\circ+\Delta \chi_{\phi-\omega}(s)$ is
	 the phase value predicted in Ref.\cite{faza}}
\label{tab5}
\end{center}
\end{table}

\begin{table}[h]
\begin{center}
\begin{tabular}[t]{cccccc}
 $N$& 1 & 2 & 3 & 4 & 5 \\
 $m_{\phi}$ & $1019.46 \pm 0.03$ & $1019.46 \pm 0.03$ & $1019.44 \pm 0.03$ &
              $1019.44 \pm 0.03$ & $1019.44 \pm 0.04$  \\
 (MeV)& & & & & \\
 $\Gamma_{\phi}$ & $4.28 \pm 0.06$ & $4.26 \pm 0.06$ & $4.22 \pm 0.08$ &
                   $4.22 \pm 0.08$ & $4.22 \pm 0.08$  \\
 (MeV)& & & & & \\ 
 $\sigma(\phi\to K^+K^-)$ & $1940 \pm 31$ & $1962 \pm 38$
                          & $1972 \pm^{63}_{58}$ & $1969 \pm 64$
			  & $1950 \pm 32$ \\
 (nb)& & & & & \\
 $\phi_{K\overline{K}}$ & 180 & $157 \pm 24$ & $156 \pm^{51}_{39}$
                        & 180 & 180 \\
 (degree)& & & & & \\
 $\mbox{Re}(A^0_{K^+K^-})$ & 0 & 0 & $4.7 \pm^{14}_{5.3}$&$7.1 \pm^{25}_{7}$ &
                           $5.9 \pm 4.8$ \\
 (MeV$^{3/2}$)& & & & & \\
 $\mbox{Im}(A^0_{K^+K^-})$ & 0 & 0 & 0 & $12 \pm 33$ & 0 \\
 (MeV$^{3/2}$)& & & & & \\
 $\chi^2/N_{df}$ & 22.15/25 & 21.2/24 & 20.45/23 & 20.54/23 & 20.65/24
 \\ \hline
\end{tabular}
\caption{Fit results for the process $e^+e^- \to K^+K^-$.
         The column number $N$ corresponds to the different variants of choice
         of the 
         phase $\phi_{K\overline{K}}$ and the real and imaginary parts of the 
         amplitude $\mbox{Re}(A^0_{K^+K^-})$,
	 $\mbox{Im}(A^0_{K^+K^-})$.}
\label{tab41}
\end{center}
\end{table}

\begin{table}[h]
\begin{center}
\begin{tabular}[t]{cccccc}
 $N$& 1 & 2 & 3 & 4 & 5 \\
 $m_{\phi}$ & $1019.40 \pm 0.02$ & $1019.38 \pm 0.02$ & $1019.38 \pm 0.02$ &
              $1019.38 \pm 0.02$ & $1019.38 \pm 0.02$ \\
 (MeV)& & & & & \\
 $\Gamma_{\phi}$ & $4.20 \pm 0.04$  & $4.15 \pm 0.04$  & $4.15 \pm 0.05$ &
                   $4.16 \pm 0.04$ & $4.16 \pm 0.04$ \\
 (MeV)& & & & & \\ 
 $\sigma(\phi\to K_SK_L)$ & $1454 \pm 12$ & $1436 \pm 12$
                          & $1450 \pm^{73}_{51}$ & $1440 \pm^{76}_{40}$
			  & $1458 \pm 12$ \\
 (nb)& & & & & \\
 $\phi_{K\overline{K}}$ & 180 & $102 \pm 14$ & $120 \pm 51$
                        & 180 & 180 \\
 (degree)& & & & & \\
 $\mbox{Re}(A^0_{K_SK_L})$ & 0 & 0 &$8 \pm^{30}_{40}$ & $4 \pm^{5}_{7}$ &
                             $5.7 \pm 2.8$ \\
 (MeV$^{3/2}$)& & & & & \\
 $\mbox{Im}(A^0_{K_SK_L})$ & 0 & 0 & 0 &$-3 \pm^{50}_{26}$ & 0  \\
 (MeV$^{3/2}$)& & & & & \\ 
 $\chi^2/N_{df}$ & 64.11/57 & 59.95/56 & 59.94/55 & 59.95/55 & 59.96/56 \\
\end{tabular}
\caption{Fit results for the process $e^+e^- \to K_SK_L$.
         The column number $N$ corresponds to the different variants of
         choice of the
         phase $\phi_{K\overline{K}}$ and 
         the real and imaginary parts of the amplitude
         $\mbox{Re}(A^0_{K_SK_L})$,
	 $\mbox{Im}(A^0_{K_SK_L})$ choice.}
\label{tab42}
\end{center}
\end{table}

\begin{table}[h]
\begin{center}
\begin{tabular}[t]{cccccc}
  $N$    & 1     & 2     & 3 & 4 & 5 \\
 $m_{\phi}-1000$ & $19.417 \pm 0.014$ & $19.419 \pm 0.014$ &
 $19.419 \pm 0.014$ &
              $19.418 \pm 0.014$ & $19.40 \pm 0.02$ \\
 (MeV) & & & & & \\
 $\Gamma_{\phi}$ & $4.21 \pm 0.04$ & $4.21 \pm 0.03$ & $4.21 \pm 0.03$ &
                   $4.21 \pm 0.03$ & $4.18 \pm 0.03$ \\
 (MeV) & & & & & \\ 
 $\sigma(\phi\to K_SK_L)$ & $1451 \pm 10$ & $1451 \pm 10 $ & $1451 \pm 10$ &
                            $1441 \pm 12$ & $1455 \pm 10$ \\
 (nb) & & & & &  \\ 
 $\sigma(\phi\to K^+K^-)$ & $1968 \pm 20$ & $1968 \pm 20$ & $1967 \pm 20$ &
                            $1994 \pm 27$ & $1961 \pm 21$ \\
 (nb) & & & & &  \\ 
 $\sigma(\phi\to 3\pi)$ & $701 \pm 6$ & $659 \pm 6$ & $664 \pm^{48}_{18}$ &
                          $702 \pm 6$ & $705 \pm 6$ \\
 (nb) & & & & & \\
 $\mbox{Re}(A^0_{K^+K^-})$ & 0 & 0 & 0 & 0 & $8.6 \pm 3.9$  \\
 (MeV$^{3/2}$) & & & & & \\ 
 $\mbox{Re}(A^0_{K_SK_L})$ & 0 & 0 & 0 & 0 & $4.3 \pm 2.5$  \\
 (MeV$^{3/2}$) & & & & & \\ 
 $\mbox{Re}(A^0_{3\pi}) \cdot 10^3$&$-46 \pm 3$&$-46 \pm 3$&$-46 \pm 3$&
                  $-46 \pm 3$&$-46 \pm 3$ \\
 (MeV$^{1/2}$) & & & & & \\ 
 $\chi_{\phi-\omega}$ & $180$&$180+\Delta\chi_{\phi-\omega}(s)$ &
 $165 \pm^{19}_{7}$
                      & $180$ & $180$ \\
 (degree) & & & & & \\
 $\phi_{K\overline{K}}$&$180$&$180$&$180$&$147 \pm^{27}_{16}$&$180$ \\
 (degree) & & & & & \\ 
  $\chi^2/N_{df}$& 110.29/114 & 110.09/114 & 110.1/113 & 108.67/113 &
  103.39/111 \\
\end{tabular}
\caption{Results of the combined fit for the processes
         $e^+e^- \to K\overline{K}$ and $3\pi$.
	 The column number $N$ corresponds to the different variants of 
         choice of the phases
	 $\phi_{K\overline{K}}$, $\chi_{\phi-\omega}$ and the real parts of
         the amplitudes
	 $\mbox{Re}(A^0_{K^+K^-})$, $\mbox{Re}(A^0_{K_SK_L})$,
	 $\mbox{Re}(A^0_{3\pi})$.}
\label{tab6}
\end{center}
\end{table}

\begin{table}[h]
\begin{center}
\begin{tabular}[t]{ccc}
  & SND & Other data \\ \hline
 $\sigma(\phi\to K_SK_L)$, nb & $1451 \pm 49 $ & $1367 \pm 26$ \cite{cmd3} \\
 $\sigma(\phi\to K^+K^-)$, nb & $1968 \pm 142$ & $2001 \pm 105$ \cite{cmd1} \\
 $\sigma(\phi\to 3\pi)$, nb & $659 \pm 35$ & $654 \pm 40$ \cite{cmd1}\\
 $B(\phi \to e^+e^-) \times 10^{4}$ & $2.93 \pm 0.14$ & $2.91 \pm 0.07$
 \cite{pdg} \\
 $B(\phi \to K^+K^-), \%$ & $47.6 \pm 1.7 $ & $49.2 \pm 0.7$ \cite{pdg} \\
 $B(\phi \to K_SK_L), \%$ & $35.1 \pm 1.3$ & $33.8 \pm 0.6$ \cite{pdg} \\
 $B(\phi \to 3\pi), \%$ & $15.9 \pm 0.8$ & $15.5 \pm 0.6$ \cite{pdg} \\ 
 $B(\phi \to \eta\gamma),\%$ & $1.33 \pm 0.06$ &  $1.297 \pm 0.033$
 \cite{pdg} \\
 $m_\phi$, MeV &$1019.42 \pm 0.05$& $1019.417 \pm 0.014$ \cite{pdg} \\
 $\Gamma_\phi$, MeV & $4.21 \pm 0.04 $& $4.458 \pm 0.032 $ \cite{pdg} \\
\end{tabular}
\caption{The comparison of the main results of this work with PDG data
         \cite{pdg} and results of the other experiments \cite{cmd1,cmd3}.}
\label{tab10}
\end{center}
\end{table}
\newpage
\begin{table}
\begin{center}
\begin{tabular}[t]{ccccc}
 $\sqrt[]{s}$&$\sigma_{K^+K^-}$&$\sigma_{K_SK_L}$&$\sigma_{K_SK_L}$&
 $\sigma_{3\pi}$ \\
             &   & (charged mode) & (neutral mode) & \\ \hline
\multicolumn{5}{c}{SCAN PHI9801} \\ \hline
  984.21 &$$&$$&&$18.1 \pm 0.9$\\
 1003.91 &$             $&$7.7 \pm 3.0$&$7.7 \pm 1.0$&$36.2 \pm 1.3$\\
 1010.17 &$48.1 \pm 13.3$&$45.2 \pm 6.1$&$34.0 \pm 2.8$&$68.5 \pm 2.4$\\ 
 1015.75 &$409.0 \pm 20.3$&$315.5 \pm 20.2$&$294.5 \pm 13.3$&$243.1 \pm 7.5$\\
 1016.68 &$717.1 \pm 32.3$&$525.5 \pm 25.7$&$513.1 \pm 20.6$&$358.9 \pm 10.6$
 \\ 
 1017.59 &$1112.0 \pm 51.8$&$836.3 \pm 37.6$&$809.1 \pm 32.5$&$493.6 \pm 14.9$
 \\
 1018.78 &$1794.1 \pm 64.8$&$1324.7 \pm 40.4$&$1295.0 \pm 33.7$&
 $658.6 \pm 11.6$\\
 1019.79 &$1898.1 \pm 56.9$&$1388.4 \pm 36.2$&$1372.4 \pm 23.9$&
 $595.5 \pm 14.1$\\ 
 1020.65 &$1471.9 \pm 54.5$&$1053.0 \pm 35.6$&$1051.9 \pm 30.3$&
 $399.8 \pm 14.5$\\
 1021.68 &$928.4 \pm 37.7$&$648.0 \pm 29.0$&$667.2 \pm 22.3$&$217.4 \pm 8.5$\\
 1023.27 &$514.4 \pm 20.0$&$352.0  \pm 16.7$&$359.5 \pm 11.7$&$92.2 \pm 3.4$\\
 1028.23 &$161.1 \pm  6.0$&$103.0 \pm 7.0$&$108.1 \pm 4.0$&
 $11.2 \pm 0.65 \pm 0.036$\\
 1033.84 &$72.9 \pm 2.9$&$45.1 \pm 4.0$&$52.0 \pm 2.1$&
 $1.400 \pm 0.113 \pm^{0.100}_{0.060}$\\
 1039.59 &$47.4 \pm 2.0$&$28.3 \pm 3.1 \pm 0.1$&$31.0 \pm 1.5 \pm 0.4$&
 $0.096 \pm 0.010 \pm^{0.09}_{0.05}$\\ 
 1049.81 &$29.0 \pm 1.5$&$13.3 \pm 2.1 \pm 0.6$&$16.4 \pm 0.9 \pm 0.5$&
 $0.613 \pm 0.081 \pm^{0.080}_{0.040}$\\ 
 1059.66 &$18.6 \pm 1.2$&$17.0 \pm 2.6 \pm 2.2$&$11.0 \pm 0.7 \pm 0.5$&
 $1.304 \pm 0.194 \pm^{0.070}_{0.040}$\\ \hline
\multicolumn{5}{c}{SCAN PHI9802} \\ \hline
 984.02  &$$&$$&&$17.3 \pm 0.8$\\
 1003.71 &$             $&$7.4 \pm 2.9$&$6.7 \pm 1.0$&$37.6  \pm 1.4$\\
 1010.34 &$38.7 \pm 12.8$&$40.0 \pm 6.3$&$33.7 \pm 2.7$&$69.5  \pm 2.5$\\ 
 1015.43 &$354.2 \pm 17.2$&$288.3 \pm 17.2$&$265.9 \pm 11.2$&$220.0 \pm 6.5$\\ 
 1016.78 &$734.7 \pm 34.2$&$564.0 \pm 27.1$&$532.6 \pm 21.7$&
 $353.6 \pm 11.1$\\ 
 1017.72 &$1181.8 \pm 54.7$&$923.1 \pm 39.5$&$875.1 \pm 34.1$&
 $515.0 \pm 15.3$\\
 1018.62 &$1726.7 \pm 65.8$&$1337.1 \pm 42.7$&$1268.5 \pm 36.3$&
 $664.2 \pm 13.1$\\ 
 1019.51 &$1946.2 \pm 56.7$&$1473.8 \pm 34.7$&$1426.3 \pm 21.6$&
 $667.0 \pm 11.8$\\ 
 1020.43 &$1639.6 \pm 56.8$&$1193.1 \pm 37.2$&$1165.2 \pm 30.6$&
 $471.2 \pm 15.5$\\
 1021.41 &$1087.2 \pm 42.2$&$757.5 \pm 32.5$&$767.0 \pm 24.7$&$270.1 \pm 9.9$\\
 1022.32 &$672.9 \pm 29.8$&$465.9 \pm 24.5$&$475.4 \pm 17.7$&$142.9 \pm 6.1$\\ 
 1027.52 &$179.6 \pm 6.9$&$123.8 \pm 7.8$&$125.3 \pm 4.2$&
 $15.803 \pm 0.752 \pm 0.037$\\
 1033.58 &$79.9 \pm 2.9$&$55.4 \pm 4.2$&$54.8 \pm 2.0$&
 $1.737 \pm 0.113  \pm^{0.100}_{0.07}$\\ 
 1039.64 &$48.7 \pm 1.9$&$27.3 \pm 3.0 \pm 0.1$&$29.6 \pm 1.4 \pm 0.4$&
 $0.094  \pm 0.009 \pm^{0.08}_{0.06}$\\ 
 1049.60 &$31.2 \pm 1.5$&$15.0 \pm 2.1 \pm 0.7$&$17.5 \pm 0.9 \pm 0.5$&
 $0.595 \pm 0.062 \pm^{0.080}_{0.04}$\\
 1059.52 &$20.9 \pm 1.2$&$14.0 \pm 2.5 \pm 1.8$&$11.7 \pm 0.7 \pm 0.5$&
 $1.238 \pm 0.151 \pm^{0.060}_{0.04}$ \\
\end{tabular}
\caption{The $e^+e^- \to K\overline{K}$ and $3\pi$ cross sections. First error
         is statistical and the second is the model uncertainty. The systematic
         errors are 7.1\%  for $e^+e^- \to K^+K^-$, 4\% and 4.2\% for
	 $e^+e^- \to K_SK_L$ at charged and neutral modes and 5\% for
	 $e^+e^- \to 3\pi$.}
\label{tab7}
\end{center}
\end{table}


 
\end{document}